\documentclass[11pt,a4paper]{article}
\pdfoutput=1

\usepackage[T1]{fontenc}

\usepackage{jheppub}
\usepackage{psfrag}
\usepackage{slashed}
\usepackage{cancel}
\usepackage{lscape}

\usepackage{caption}
\usepackage{array}
\usepackage{graphicx}
\usepackage{subcaption}
\usepackage{multirow}

\usepackage[utf8]{inputenc}

\definecolor{mygreen}{rgb}{0,0.7,0}

\def\la{\langle}
\def\ra{\rangle}
\def\spA#1#2{\la#1#2\ra}
\def\spB#1#2{[#1#2]}
\def\spAB#1#2#3{\la#1|#2|#3]}

\def\spAA#1#2#3{\la#1|#2|#3\ra}

\def\spab#1#2{\la#1|#2]}
\def\spaa#1#2#3#4{\la#1#2#3#4\ra}

\DeclareMathOperator{\tr}{\rm tr}

\def\eps{\epsilon}

\def\tb{{\bar{t}}}

\def\s#1{s_{#1}}
\def\d#1#2{#1\cdot #2}
\def\p#1{#1}
\def\pp#1{p_{#1}}
\def\f#1{#1^\flat}
\def\n#1{\eta_{#1}}

\def\Adcc{B_n^{(1)}}

\def\usepic#1#2{\parbox{#1}{\includegraphics[width=#1]{#2}}}

\preprint{{Edinburgh 2017/07, IPPP/17/22}}

\title{A unitarity compatible approach to one-loop amplitudes with massive fermions}

\author[a]{Simon Badger,}
\author[b]{Christian Br\o{}nnum-Hansen,}
\author[a]{Francesco Buciuni,}
\author[b]{Donal O'Connell}
\affiliation[a]{
Institute for Particle Physics Phenomenology, Department of Physics, Durham University, Durham DH1
3LE, United Kingdom%
}
\affiliation[b]{
Higgs Centre for Theoretical Physics, School of Physics and Astronomy, The University of Edinburgh, Edinburgh EH9 3JZ, Scotland, UK%
}
\emailAdd{simon.d.badger@durham.ac.uk, bronnum.hansen@ed.ac.uk, francesco.buciuni@durham.ac.uk,
donal@staffmail.ed.ac.uk}

\abstract{We explain how one-loop amplitudes with massive fermions can be computed using only on-shell
information. We first use the spinor-helicity formalism in six dimensions to perform generalised
unitarity cuts in $d$ dimensions. We then show that divergent wavefunction cuts can be avoided,
and the remaining ambiguities in the renormalised amplitudes can be fixed, by matching to universal infrared poles in $4-2\eps$
dimensions and ultraviolet poles in $6-2\eps$ dimensions. In the latter case we construct an effective
Lagrangian in six dimensions and reduce the additional constraint to an on-shell tree-level
computation.}

\begin{document}
\maketitle
\flushbottom

\section{Introduction \label{sec:intro}}

Precise predictions for the production of strongly interacting massive particles
are in high demand for current experimental analyses at the LHC. The current
precision level of predictions is in relatively good shape, with top quark pair production now known
differentially at NNLO in QCD~\cite{Czakon:2013goa,Czakon:2015owf} and a full range of off-shell decays known at NLO in QCD with an
additional jet~\cite{Bevilacqua:2015qha}.
Modern one-loop techniques are also able to explore high multiplicity final states where
the current state-of-the-art is top quark pair production in association with three
jets~\cite{Hoche:2016elu}. The GoSam collaboration has also been able to produce NLO
predictions for the challenging $t\tb H+j$ final state~\cite{vanDeurzen:2013xla}. A more complete
overview of the current status can be found in reference~\cite{Badger:2016bpw}.

On the other hand, these processes are often overlooked by more formal studies of amplitudes in gauge
theory which can uncover hidden simplicity and structure. While it is well known that on-shell
techniques like unitarity \cite{Bern:1995db}, spinor integration~\cite{Britto:2006fc,Britto:2008vq} and BCFW recursion apply equally well to massive
amplitudes, explicit computations are relatively few~\cite{Rozowsky:1997dm,Badger:2010mg,Badger:2011yu}. Nevertheless some computations using
these approaches have produced compact analytic results useful for phenomenological applications
~\cite{Badger:2010mg,Campbell:2012uf}. While elements of these computations use unitarity cuts and
on-shell trees, Feynman diagrams techniques were also employed to compute the UV counterterms
necessary for mass and wavefunction renormalisation. To the best of our knowledge the only computations not to do
this are those with a massive internal loop where a UV matching prescription was used
\cite{Bern:1995db,Rozowsky:1997dm}.

\begin{figure}[b]
  \begin{center}
    \includegraphics[width=0.8\textwidth]{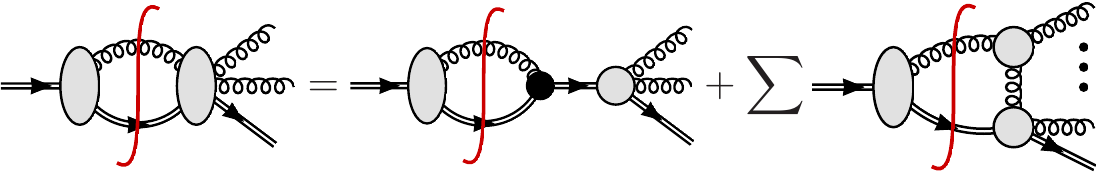}
  \end{center}
  \caption{Decomposing the tree amplitude appearing on the left hand side of a wavefunction
  cut reveals a divergent graph.}
  \label{fig:wfcut}
\end{figure}

The obstacle is that the traditional approach to renormalisation requires the
amputation of wavefunction graphs, and the addition of counterterm diagrams. This procedure breaks
gauge invariance during intermediate steps and therefore causes problems for methods based on
(generalised) unitarity~\cite{Bern:1994zx,Bern:1994cg,Britto:2004nc}, which construct amplitudes from on-shell tree-level building
blocks. Naive attempts to amputate wavefunction graphs 
in generalised unitarity are precluded by the presence of an on-shell propagator, leading to a factor $1/0$: this is depicted explicitly in figure \ref{fig:wfcut},
where the on-shell tree amplitude appearing on the right hand side of a two-particle cut is expanded
to reveal a divergent propagator inside. Consequently, the favoured method is still to follow an approach based on Feynman diagrams; then the
amputation of wavefunction graphs is straightforward. 

Two solutions to this problem have been proposed. Ellis, Giele, Kunszt and Melnikov showed that
modifying the tree-level input entering the double cuts of the wavefunction graphs allowed a simple
implementation of the on-shell renormalisation scheme~\cite{Ellis:2008ir}. All cuts can then be
performed but gauge invariance is only restored at the end of the computation.  Since the removal of
the unwanted graphs is extremely easy to implement within a Berends-Giele construction of the
tree-level amplitudes in the cuts this method is quite efficient numerically.  A second solution,
proposed by Britto and Mirabella \cite{Britto:2011cr}, is to regulate the divergent tree by
introducing a momentum shift. This procedure allows us to preserve gauge invariance but introduces
an additional variable into the calculation which will cancel when combined with the
mass-renormalisation counterterms. In either case a set of extra two- and single-particle cuts is
necessary together with the counterterms to fully determine the amplitude in comparison to the
massless case.

Despite both of these solutions there is still an open question: is it possible to compute
amplitudes with masses using only on-shell gauge invariant building blocks and without introducing
additional regulators? Both of the approaches mentioned above follow the on-shell
renormalisation scheme where divergences can be absorbed into additional terms in the Lagrangian.
In this paper we will seek an alternative way to absorb the divergences by appealing to an
effective six dimensional version of QCD.

This procedure relies on first computing a full set of finite $d$-dimensional unitarity cuts.
We show how this can be done efficiently in the six-dimensional spinor-helicity
formalism~\cite{Cheung:2009dc} by embedding the additional mass into the higher dimensions and performing cuts
in six dimensions. In particular we show how these results can be dimensionally reduced to
$d$-dimensional amplitudes keeping the spin dimension of the gluon $d_s$ arbitrary. This
generalises the previous approaches used for massless cuts in six-dimensions~\cite{Bern:2010qa,Davies:2011vt}.

Our paper is organised as follows. We begin by reviewing the structure of one-loop amplitudes in
dimensional regularisation and the integrand level representations. In section \ref{sec:massivespinors}
we review the spinor-helicity formalisms in four and six dimensions, and
show how Dirac spinors for massive fermions can be represented as massless Weyl spinors in six
dimensions. We then discuss a simple example of a pair of massive fermions coupling to an off-shell vertex
at one-loop. This example allows us to show how computations in six dimensions can be performed, and how they can be
dimensionally reduced to results with an arbitrary spin dimension $d_s$. In section \ref{sec:6Dcuts}
we summarise the generalised unitarity method in six dimensions and explain some of the key features
needed to apply it to the case of $gg\to t\tb$ scattering. Section \ref{sec:wftadcoeff} describes the
procedure of fixing the remaining ambiguities using the universal epsilon pole structure in $d = 4-2\eps$ dimensions
and the corresponding epsilon pole structure of the effective theory in $6-2\eps$ dimensions. Following our conclusions, the appendices
give further details of the spinor-helicity method, the tree-level amplitudes in six dimensions
and multiple cut solutions. We also include an appendix with the Feynman rules for the
dimension six operators used in section \ref{sec:wftadcoeff}.

In addition we provide a \textsc{Mathematica} workbook with the arXiv version of this article. This
workbook contains a basic implementation of all the ingredients used in the paper and runs through
numerical examples for two leading colour primitive one-loop amplitudes contributing to $gg\to t\tb$
scattering. Analytic formulae for these two primitive amplitudes are also provided in a computer
readable form including the full dependence on the loop momentum dimension $d$ and spin dimension $d_s$.

\section{Review of one-loop amplitudes and integrands \label{sec:1lsummary}}

In this section we review the basics of one-loop integrand parametrisations in $d$ dimensions.
While many good reviews of the subject are available, e.g.
\cite{Bern:2007dw,Dixon:2013uaa,Ellis:2011cr}, this section also introduces the notation we will use
in the rest of the paper. The amplitudes we will consider in this paper are QCD amplitudes with one massive
fermion flavour. In this case there are only two possible basis integrals which go beyond those appearing in the
massless case,
\begin{align}
  A_n^{(1)} = \Adcc + c_{2;m^2} I_{2,m^2} + c_1 I_1.
  \label{eq:1lampdecomp}
\end{align}
The amplitude labelled $\Adcc$ is the part that can be constructed from finite $d$-dimensional
unitarity cuts. The additional basis integrals depend only on the fermion mass and in dimensional regularisation are,
\begin{align}
  I_{2,m^2}  &= \mu_R^{2\epsilon}
      \int \frac{d^dk}{(2\pi)^d} \frac{1}{k^2((k-p)^2-m^2)}
      \overset{d=4-2\eps}{=} i c_\Gamma \left(\frac{1}{\eps} + \log\left( \frac{\mu_R^2}{m^2} \right)
      + 2\right) + \mathcal{O}(\eps),
  \label{eq:wfint}\\
  I_1 &= \mu_R^{2\epsilon}
      \int \frac{d^dk}{(2\pi)^d} \frac{1}{k^2-m^2}
      \overset{d=4-2\eps}{=} i c_\Gamma m^2 \left( \frac{1}{\eps} + \log\left( \frac{\mu_R^2}{m^2} \right) + 1 \right) + \mathcal{O}(\eps),
  \label{eq:tadint}
\end{align}
where $c_\Gamma = \frac{\Gamma(1+\eps)\Gamma(1-\eps)^2}{(4\pi)^{2-\eps} \Gamma(1-2\eps)}$.

The amplitudes $\Adcc$ can be written in the usual integrand basis
of irreducible scalar products including extra dimensional terms following the
OPP~\cite{Ossola:2006us}/EGKM~\cite{Ellis:2007br,Giele:2008ve} constructions,
\begin{align}
  \Adcc =& \mu_R^{2\epsilon} \int \frac{d^dk}{(2\pi)^d} \Bigg\{\nonumber\\&
    \sum_{1\leq i_1<i_2<i_3<i_4<i_5\leq n} \frac{\Delta_{\{i_1,i_2,i_3,i_4,i_5\}}}{D_{i_1} D_{i_2} D_{i_3} D_{i_4} D_{i_5}}
  + \sum_{1\leq i_1<i_2<i_3<i_4\leq n} \frac{\Delta_{\{i_1,i_2,i_3,i_4\}}}{D_{i_1} D_{i_2} D_{i_3} D_{i_4}}
    \nonumber\\&
  + \sum_{1\leq i_1<i_2<i_3\leq n} \frac{\Delta_{\{i_1,i_2,i_3\}}}{D_{i_1} D_{i_2} D_{i_3} D_{i_4}}
  + \sum_{\substack{1\leq i_1<i_2\leq n \\ i_2-i_1 \text{ mod } n > 1}} \frac{\Delta_{\{i_1,i_2\}}}{D_{i_1} D_{i_2} D_{i_3} D_{i_4}}
  \Bigg\}.
  \label{eq:6dintegrandbasis}
\end{align}
For renormalisable gauge theories we can give a complete parametrisation of the numerators,
\begin{subequations}
\begin{align}
  \Delta_{\{i_1,i_2,i_3,i_4,i_5\}} &=
      c^{(0)}_{i_1,i_2,i_3,i_4,i_5}\,  \mu^2 , \\
  \Delta_{\{i_1,i_2,i_3,i_4\}} &=
      c^{(0)}_{i_1,i_2,i_3,i_4}
    + c^{(1)}_{i_1,i_2,i_3,i_4}\,  (k\cdot w_{1;i_1,i_2,i_3})
    + c^{(2)}_{i_1,i_2,i_3,i_4}\,  \mu^2 \nonumber\\&
    + c^{(3)}_{i_1,i_2,i_3,i_4}\,  \mu^2 (k\cdot w_{1;i_1,i_2,i_3})
    + c^{(4)}_{i_1,i_2,i_3,i_4}\,  \mu^4 , \\
  \Delta_{\{i_1,i_2,i_3\}} &=
      c^{(0)}_{i_1,i_2,i_3}
    + c^{(1)}_{i_1,i_2,i_3}\,  (k\cdot w_{1;i_1,i_2})
    + c^{(2)}_{i_1,i_2,i_3}\,  (k\cdot w_{2;i_1,i_2}) \nonumber\\&
    + c^{(3)}_{i_1,i_2,i_3}\,  (k\cdot w_{1;i_1,i_2}) (k\cdot w_{2;i_1,i_2})
    + c^{(4)}_{i_1,i_2,i_3}\,  \left((k\cdot w_{1;i_1,i_2})^2 -  (k\cdot w_{2;i_1,i_2})^2\right) \nonumber\\&
    + c^{(5)}_{i_1,i_2,i_3}\,  (k\cdot w_{1;i_1,i_2})^2 (k\cdot w_{2;i_1,i_2})
    + c^{(6)}_{i_1,i_2,i_3}\,  (k\cdot w_{1;i_1,i_2}) (k\cdot w_{2;i_1,i_2})^2 \nonumber\\&
    + c^{(7)}_{i_1,i_2,i_3}\,  \mu^2 (k\cdot w_{1;i_1,i_2})
    + c^{(8)}_{i_1,i_2,i_3}\,  \mu^2 (k\cdot w_{2;i_1,i_2})
    + c^{(9)}_{i_1,i_2,i_3}\,  \mu^2, \\
  \Delta_{\{i_1,i_2\}} &=
      c^{(0)}_{i_1,i_2}
    + c^{(1)}_{i_1,i_2}\,  (k\cdot w_{1;i_1})
    + c^{(2)}_{i_1,i_2}\,  (k\cdot w_{2;i_1})
    + c^{(3)}_{i_1,i_2}\,  (k\cdot w_{3;i_1}) \nonumber\\&
    + c^{(4)}_{i_1,i_2}\,  (k\cdot w_{1;i_1}) (k\cdot w_{2;i_1})
    + c^{(5)}_{i_1,i_2}\,  (k\cdot w_{1;i_1}) (k\cdot w_{3;i_1}) \nonumber\\&
    + c^{(6)}_{i_1,i_2}\,  (k\cdot w_{2;i_1}) (k\cdot w_{3;i_1})
    + c^{(7)}_{i_1,i_2}\,  \left((k\cdot w_{1;i_1})^2  - (k\cdot w_{3;i_1})^2\right) \nonumber\\&
    + c^{(8)}_{i_1,i_2}\,  \left((k\cdot w_{2;i_1})^2  - (k\cdot w_{3;i_1})^2\right)
    + c^{(9)}_{i_1,i_2}\,  \mu^2.
\end{align}
\label{eq:deltaparam}%
\end{subequations}
The irreducible numerators $k\cdot w_{x;i_1,\ldots,i_s}$ can be constructed using the spurious directions of
van Neerven and Vermaseren~\cite{vanNeerven:1983vr} and vanish after integration. The spurious
directions $w_{x;i_1,\ldots,i_p}$ are orthogonal to the $p$ dimensional physical space spanned by
the momenta entering vertices $i_1,\ldots,i_s$ where $x=1,\ldots,s$ with $s+p=4$. $\mu^2=-\tilde{k}\cdot\tilde{k}$ is the
extra dimensional scalar product where $k = \bar{k} + \tilde{k}$.
These give rise to dimension shifted integrals which in turn lead to rational terms in $d=4-2\eps$ dimensions.
The coefficients $c$ are rational functions of the external kinematics and can be extracted from
generalised unitarity cuts. After elimination of vanishing integrals over the spurious directions,
the $d$-dimensional representation of the amplitude is,
\begin{align}
  &\Adcc(d,d_s) =
    \sum_{1\leq i_1<i_2<i_3<i_4<i_5\leq n}
        c^{(0)}_{i_1,i_2,i_3,i_4,i_5}\,  I^d_{i_1,i_2,i_3,i_4,i_5}[\mu^2]
    \nonumber\\&
  + \sum_{1\leq i_1<i_2<i_3<i_4\leq n}
        c^{(0)}_{i_1,i_2,i_3,i_4}\,  I^d_{i_1,i_2,i_3,i_4}[1]
      + c^{(2)}_{i_1,i_2,i_3,i_4}\,  I^d_{i_1,i_2,i_3,i_4}[\mu^2]
      + c^{(4)}_{i_1,i_2,i_3,i_4}\,  I^d_{i_1,i_2,i_3,i_4}[\mu^4]
    \nonumber\\&
  + \sum_{1\leq i_1<i_2<i_3\leq n}
        c^{(0)}_{i_1,i_2,i_3}\,  I^d_{i_1,i_2,i_3}[1]
      + c^{(9)}_{i_1,i_2,i_3}\,  I^d_{i_1,i_2,i_3}[\mu^2]
    \nonumber\\&
  + \sum_{\substack{1\leq i_1<i_2\leq n \\ i_2-i_1 \text{ mod } n > 1}}
        c^{(0)}_{i_1,i_2}\,  I^d_{i_1,i_2}[1]
      + c^{(9)}_{i_1,i_2}\,  I^d_{i_1,i_2}[\mu^2],
  \label{eq:6dintegralbasis}
\end{align}
where
\begin{align}
  I^d_{i_1,i_2,\ldots,i_n}[N] = \mu_R^{2\epsilon} \int \frac{d^dk}{(2\pi)^d} \frac{N}{D_{i_1} D_{i_2} \cdots D_{i_n}}.
  \label{eq:integraldef}
\end{align}
Explicitly the dimension shifting relations are \footnote{A simple derivation of this fact is shown in appendix A.2 of reference~\cite{Bern:1995db}},
\begin{align}
  I^d_{i_1,i_2,\ldots,i_n}[\mu^2] &= \frac{d-4}{2} (4\pi) 
   I^{d+2}_{i_1,i_2,\ldots,i_n}[1], \\
  I^d_{i_1,i_2,\ldots,i_n}[\mu^4] &= \frac{(d-4)(d-2)}{4} (4\pi)^2 
  I^{d+4}_{i_1,i_2,\ldots,i_n}[1].
  \label{eq:dimshift}
\end{align}

\section{Massive fermion spinors  \label{sec:massivespinors}}

In this section we describe how we can use massless six dimensional momenta to obtain amplitudes in four dimensions with massive particles. Before getting started it is helpful to recall how massive fermion wavefunctions
can be incorporated within the four-dimensional spinor-helicity formalism commonly used for
massless amplitudes. Here we follow the notation used previously in reference \cite{Badger:2011yu}
while the formalism itself has been well established long before that, see for example
\cite{Kleiss:1985yh,Hagiwara:1985yu,Schwinn:2005pi,Rodrigo:2005eu,Maitre:2007jq}.

Starting from a massive 4-momentum $p$ with $p^2=m^2$, we can define a massless projection
with respect to a light-like reference vector $\eta$,
\begin{align}
  p^\flat = p - \frac{m^2}{2 p\cdot \eta} \eta,
  \label{eq:pflatdef}
\end{align}
such that $(p^\flat)^2 = 0$. The massive spinors can then be constructed from the Weyl spinors of
$p^\flat$ and $\eta$,
\begin{align}
  \bar{u}_+(p,m) &= \frac{\la \eta |(\slashed{p}+m)}{\spA{\eta}{\f{p}}}, &
  \bar{u}_-(p,m) &= \frac{[\eta|(\slashed{p}+m)}{\spB{\eta}{\f{p}}}, \nonumber\\
  v_+(p,m) &= \frac{(\slashed{p}-m)|\eta\ra}{\spA{\f{p}}{\eta}}, &
  v_-(p,m) &= \frac{(\slashed{p}-m)|\eta]}{\spB{\f{p}}{\eta}}.
  \label{eq:4dspinordef}
\end{align}
Tree-level helicity amplitudes for $gg\to t\tb$ scattering can then be written in a relatively
compact notation using four-dimensional spinor products,
\begin{align}
  -i&\spA{\n1}{\f1}\spA{\n4}{\f4} A^{(0)}(1_t^+,2^+,3^+,4_\tb^+) =
  -\frac{m_t^3 \s{23} \spA{\n{1}}{\n{4}}}{2 \d{\pp1}{\pp2}\spA{\p{2}}{\p{3}}^2},\label{eq:4dallplus} \\
  -i&\spA{\n1}{\f1}\spA{\n4}{\f4} A^{(0)}(1_t^+,2^+,3^-,4_\tb^+) =
  \frac{m_t\spA{\n{1}}{\p{3}}\spA{\n{4}}{\p{3}}\spAB{\p{3}}{\p{1}}{\p{2}}}{2\d{\pp1}{\pp2}\spA{\p{2}}{\p{3}}}
  + \frac{m_t\spA{\n{1}}{\n{4}}\spAB{\p{3}}{\p{1}}{\p{2}}^2}{2\d{\pp1}{\pp2}\s{23}}.\label{eq:4dsingleminus}
\end{align}
We will now show how these results can be rewritten in terms of amplitudes of massless fermions
in six dimensions.

\subsection{Massive fermions from massless six dimensional spinors \label{sec:6Dspinors}}

We will in the following be using the six dimensional spinor-helicity formalism
introduced in~\cite{Cheung:2009dc} and further discussed in~\cite{Boels:2009bv,Dennen:2009vk,Bern:2010qa,CaronHuot:2010rj,Dennen:2010dh,Davies:2011vt,Boels:2012ie}. However, we start by looking at a free massive fermion field in four dimensions,
\begin{align}
\mathcal{L}^{4d} &= \overline{\psi}(x) (i \gamma^\mu \partial_\mu - m) \psi(x).
\end{align}
We use the Weyl representation for the Dirac $\gamma$ matrices
\begin{align}
\gamma^\mu = \begin{pmatrix}
0 & \tilde{\sigma}^\mu \\
\sigma^\mu & 0
\end{pmatrix},
\end{align}
where $\sigma^\mu = (1, \sigma^i)$ and $\tilde{\sigma}^\mu = (1, -\sigma^i)$ and the Pauli matrices,
$\sigma^i,\ i=1,2,3$, are given in appendix \ref{app:matrices}. For the spinors associated with
external fermions we seek solutions to the massive Dirac equation
\begin{align}
(\gamma \cdot \bar{p} - m) u_s(\bar{p}) = 0\ \text{and} \ \bar{u}_s(\bar{p})(\gamma \cdot \bar{p} - m) = 0,
\end{align}
where the bar denotes that the momentum is in four dimensions. Alternatively we can consider a
massless fermion field in six dimensions,
\begin{align}
\mathcal{L}^{6d} &= \overline{\Psi} (x) (i\Gamma^M \partial_M) \Psi(x).
\end{align}
Note that for six dimensions we use capital Greek letters and $M$ runs from 0 to 5. In six
space-time dimensions the Dirac matrices are $8\times8$ objects, and we represent them by
\begin{align}
\Gamma^M = \begin{pmatrix}
0 & \tilde{\Sigma}^M \\
\Sigma^M & 0
\end{pmatrix},\label{Gamma-matrices}
\end{align}
where the $\Sigma$-matrices are defined by taking outer products of Pauli matrices and are listed
explicitly in
appendix \ref{app:matrices}. This representation gives a simple relation to the four dimensional
$\gamma$-matrices for the first four of the $\Sigma$-matrices
\begin{align}
-\tilde{\Sigma}^{5,AX}\Sigma^\mu_{XB} &= (\gamma^\mu)^A_{\ B} = \tilde{\Sigma}^{\mu,AX}\Sigma^5_{XB},\label{eq:gammarelation}
\end{align}
where we have adopted the convention that $\Sigma^M$ carry lower spinor indices while
$\tilde{\Sigma}^M$ carry upper. For the remaining two $\Sigma$-matrices we have
\begin{align}
-\tilde{\Sigma}^{5,AX} \Sigma^4_{XB} &= (-\gamma^0\gamma^1\gamma^2\gamma^3)^A_{\ B} = i (\gamma^5)^A_{\ B}, \\
-\tilde{\Sigma}^{5,AX}\Sigma^5_{XB} &= \boldsymbol{1}^A_{\ B}.
\end{align}

Since there in six dimensions is no mass term and we use this Weyl-like representation for the
$\Gamma$-matrices \eqref{Gamma-matrices}, we can decompose $\Psi=(\Psi_1, \Psi_2)$ and see that the
two fields decouple:
\begin{align}
\mathcal{L}^{6d} &= \overline{\Psi}_1(x) (i \Sigma^M \partial_M) \Psi_1(x) +\overline{\Psi}_2(x) (i \tilde{\Sigma}^M \partial_M) \Psi_2(x).\label{decoupling}
\end{align}
Hence the fields are basically copies of each other. The spinors associated with these fields have
four components and the Dirac equation in momentum space associated with $\Psi_1(x)$ reads
\begin{align}
(\Sigma \cdot p)_{AB} \lambda^B_a(p) = 0\label{eq:dirac1},
\end{align}
where the $SU(4)$ indices $A,B$ run from $1$ to $4$, and the lower case index, $a$, is a helicity
index taking two values. Hence we can regard the six dimensional spinors as $4 \times 2$ matrices.
For the spinors associated with $\Psi_2(x)$ the corresponding equation is
\begin{align}
(\tilde{\Sigma} \cdot p)^{AB} \tilde{\lambda}_{B\dot{a}}(p) = 0.
\end{align}

For massive four dimensional momentum $\bar{p}$ we choose to write six dimensional massless momentum
\begin{align}
p=(\bar{p},0,m),\ \text{so}\ p^2 = \bar{p}^2-m^2 = 0.
\label{eq:6dfor4dmass}
\end{align}
With this choice we relate the chiral $\lambda$'s to the anti-chiral $\tilde{\lambda}$'s using the identity
\begin{align}
\lambda^A = i \tilde{\Sigma}^{4,AB} \tilde{\lambda}_B.\label{eq:chiral-antichiral}
\end{align}
This can be verified by insertion and use of the Clifford algebra in the Dirac equation \eqref{eq:dirac1}:
\begin{align}
0 &= (\Sigma \cdot p)_{AB} \lambda^B_a(p)\nonumber \\
&= i (\Sigma \cdot p)_{AB} \tilde{\Sigma}^{4,BC} \tilde{\lambda}_C \nonumber \\
&= - i \Sigma^{4}_{AB} (\tilde{\Sigma} \cdot p)^{BC}  \tilde{\lambda}_C \nonumber \\
\Rightarrow 0 &= (\tilde{\Sigma} \cdot p)^{BC}  \tilde{\lambda}_C.
\end{align}
With this choice of momentum embedding the Dirac equation (\ref{eq:dirac1}) becomes
\begin{align}
(\Sigma \cdot p)_{AB} \lambda^B_a(p) &= \left(\Sigma^\mu p_\mu - \Sigma^5 p^{(5)}\right)_{AB} \lambda^B_a(p) = 0.\label{eq:dirac2}
\end{align}
By multiplying from the left by $-\tilde{\Sigma}^{5,XA}$ we obtain
\begin{align}
(\gamma \cdot \bar{p} - p_1^{(5)})^X_{\ B} \lambda^B(p) = 0.
\end{align}
Notice how the sign on the sixth component of momentum determines whether $\lambda(p)$ should be associated with the four dimensional spinor for a fermion $u(p)$ or an anti-fermion $v(p)$:
\begin{align}
\lambda (p) = \begin{cases} 
u(\bar{p}) & ,\ p^{(5)} = m \\
v(\bar{p}) & ,\ p^{(5)} = -m
\end{cases}.
\end{align}
 A similar calculation shows how to identify the six dimensional spinors with the adjoint four dimensional spinors:
\begin{align}
0 &= \lambda^A(p)(\Sigma^\mu p_\mu - \Sigma^5 p^{(5)})_{AB} \nonumber \\
&= \lambda^A(p) (-\Sigma^5 \tilde{\Sigma}^5)_A^{\ X} (\Sigma^\mu p_\mu - \Sigma^5 p^{(5)})_{XB} \nonumber \\
&= \lambda^A(p) \Sigma^5_{AX}(\gamma \cdot \bar{p} - p^{(5)})^X_{\ B}.
\end{align}
Again the sixth momentum component determines whether $\lambda (p) \Sigma^5$ should be identified with $\bar{u}(p)$ or $\bar{v}(p)$:
\begin{align}
\lambda (p) \Sigma^5 = \begin{cases} 
\bar{u}(\bar{p}) & ,\ p^{(5)} = m \\
\bar{v}(\bar{p}) & ,\ p^{(5)} = -m
\end{cases}.
\end{align}
We write an explicit representation for $\lambda^A(p)$ that allows us to make the connection with the four dimensional representation \eqref{eq:4dspinordef}.
We use a massless (in the four dimensional sense) reference vector $\eta$, as introduced in \eqref{eq:pflatdef}, with Weyl spinors $\kappa_\alpha(\eta), \tilde{\kappa}^{\dot{\alpha}}(\eta)$
and define the six dimensional spinors:
\begin{align}
\lambda^{Aa} (\eta,\bar{p}^\flat) = \begin{pmatrix}
0 & \frac{\tilde{\kappa}^{\dot{\alpha}}(\eta)}{\spB{p^\flat}{\eta}} \\
\frac{\kappa_\alpha(\eta)}{\spA{p^\flat}{\eta}} & 0
\end{pmatrix},\ \tilde{\lambda}_{A\dot{a}} (\eta,\bar{p}^\flat) = \begin{pmatrix}
0 & \frac{\tilde{\kappa}_{\dot{\alpha}}(\eta)}{\spB{p^\flat}{\eta}} \\
\frac{\kappa^\alpha(\eta)}{\spA{p^\flat}{\eta}} & 0
\end{pmatrix}.
\end{align}
We construct the massless six dimensional momentum $p=(\bar{p},0,p^{(5)})$, where $p^{(5)} = \pm m$ as discussed above. Using $ (\Sigma \cdot p)_{AB} (\tilde{\Sigma} \cdot p)^{BC} = 0$ we see that the Dirac equation (\ref{eq:dirac1}) is solved by setting
\begin{align}
\lambda^A(p) = (\tilde{\Sigma} \cdot p)^{AB}  \tilde{\lambda}_{B}(\eta,\bar{p}^\flat).
\label{eq:lambda6d}
\end{align}
The anti-chiral spinor is defined similarly:
\begin{align}
\tilde{\lambda}_A(p) = (\Sigma \cdot p)_{AB}  \lambda^{B}(\eta,\bar{p}^\flat).
\label{eq:lambdatilde6d}
\end{align}
The discussion following (\ref{eq:dirac2}) showed how these solve the massive Dirac equation in four dimensions by appropriate choice of sign for $p^{(5)}$.

Since the helicity indices for external particles are not contracted in the six dimensional spinor-helicity formalism, we obtain all helicity configurations in one amplitude. In six dimensions gluons have four polarisation states denoted by two helicity indices. The polarisation vector for a gluon with momentum $p$ is given by
\begin{align}
	\varepsilon^M_{a \dot{a}} = \frac{1}{\sqrt{2}} \spAA{p_a}{\Sigma^M}{q_b} (\spab{q_b}{p^{\dot{a}}})^{-1},
\end{align}
where $q$ is a reference vector satisfying $p \cdot q \neq 0$. Because we use Weyl spinors the fermions still have two states. The amplitude for a quark pair and two gluons is given by \cite{Davies:2011vt}
\begin{align}
	A(1_{q,a}, 2_{b \dot{b}}, 3_{c \dot{c}}, 4_{\bar{q},d}) &= \frac{-i}{2 s_{12} s_{23}} \langle 1_a 2_b 3_c 4_d \rangle [ 1_{\dot{x}} 2_{\dot{b}} 3_{\dot{c}} 1^{\dot{x}}].
\end{align}
The results in four dimensions are obtained by picking out helicity configurations $(a,b\dot{b},c\dot{c},d) = (2,22,22,2),\ (2,11,22,2)$ for the all plus \eqref{eq:4dallplus} and single minus \eqref{eq:4dsingleminus} configurations respectively. Other relevant six dimensional trees are given in appendix \ref{app:treelevel6d}.

\subsection{Interactions and state-sum reduction \label{sec:statesum}}

We introduce interactions as always by replacing the derivative with the covariant derivative. In six dimensions
\begin{align}
\partial_M \to D_M = \partial_M -igA_M^i(x) t^i,
\end{align}
where $A_M^i(x)$ are the gauge fields and $t^i$ are the generators of the gauge group. We use the decomposition
\begin{align}
A_M(x) = A_M^i(x) t^i = (A_\mu(x), \phi_1(x), \phi_2(x)),
\end{align}
which leads to the following interaction terms for $\Psi_1$ (dropping dependence on position for simplicity):
\begin{align}
\mathcal{L}^{6d}_{\text{int}, \Psi_1} &= -ig \overline{\Psi}_1 \Sigma^M A_M \Psi_1 \nonumber \\
&= -ig \overline{\Psi}_1 \left(\Sigma^\mu A_\mu - \Sigma^4 \phi_1 - \Sigma^5 \phi_2 \right) \Psi_1 \nonumber \\
&= -ig \overline{\Psi}_1 \Sigma^\mu A_\mu \Psi_1 + g \overline{\Psi}_1 \phi_1 \Psi_2 - ig \overline{\Psi}_1 \phi_2 \gamma_5 \Psi_2,\label{eq:6dlagrangian}
\end{align}
where we in the last line used the relation between chiral and anti-chiral spinors \eqref{eq:chiral-antichiral}, which for the fields reads $\Psi_1 = i \tilde{\Sigma}^{4} \Psi_2$.
The last two terms give rise to the three-point amplitudes given in (\ref{eq:amp3scalar1}) and (\ref{eq:amp3scalar5}). While the first term resembles the four dimensional interaction term the two last terms are additional contributions arising from the extra momentum components.
For internal lines these contributions correspond to additional gluon polarisation states that should be subtracted to obtain the four dimensional result. This procedure is known as state-sum reduction.

The contraction of Lorentz indices over internal propagators leads to explicit dependence on $d_s$.
Working explicitly in six dimensions this dependence will be lost but can be recovered through
state-sum reduction. The general procedure is described in \cite{Giele:2008ve,Davies:2011vt}. Gluons
have $d-2$ polarisation states, so for each extra dimension introduced we get one more state. Each
of these states correspond to the contribution from replacing gluons in the loop by a scalar. By
subtracting these scalars the number of polarisation states can be reduced to $d_s - 2$. In our
set-up, the scalar associated with the mass direction should be subtracted separately and we arrive
at the state-sum reduction prescription
\begin{align}
\boldsymbol{c} = \boldsymbol{c}^{6d} - (5 - d_s) \boldsymbol{c}_{\phi_1} - \boldsymbol{c}_{\phi_2}.
\label{eq:dimred}
\end{align}
We end this section by giving an example.

\subsection{A one-loop example calculation}

\begin{figure}
	\centering
	\includegraphics[width=0.32\linewidth]{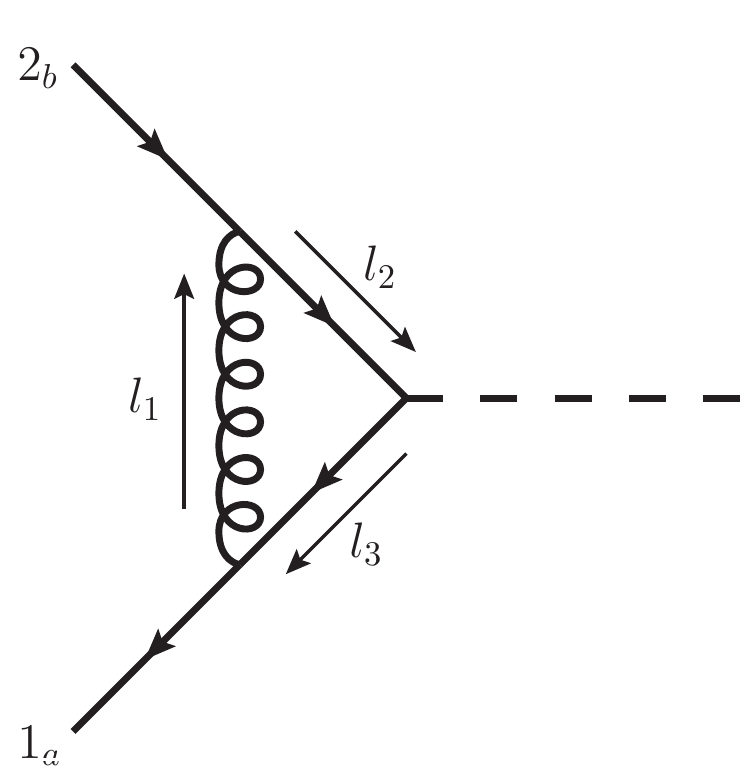}
	\caption{Feynman diagram for one-loop contribution to the coupling between a massive fermion pair and an off-shell scalar. All external momenta are outgoing.}
	\label{fig:htt}
\end{figure}

In this section we give an example of the correspondence established above by calculating the one-loop amplitude for a massive fermion pair coupling to an off-shell scalar, $A^{(1)}$. This calculation involves only one Feynman diagram (figure \ref{fig:htt}), which by using colour-ordered four dimensional Feynman rules is given by
\begin{align}
A^{(1),4d} = \int \frac{d^d \ell_1}{(2 \pi)^{d}}\ \bar{u}_1 \gamma^\mu \frac{(\gamma \cdot \ell_3 + m)}{\ell_3^2-m^2} \frac{(\gamma \cdot \ell_2 + m)}{\ell_2^2-m^2} \gamma^\nu v_2 \frac{\eta_{\mu \nu}}{\ell_1^2}
\equiv \int \frac{d^d\ell_1}{(2\pi)^{d}} \frac{N^{4d}}{D_1D_2D_3},\label{eq:feynmancalc}
\end{align}
where $\ell_2 = \ell_1-p_2$, $\ell_3=\ell_1+p_1$, $D_i = \ell_i^2 - m_i^2$, and $N^{4d}$ is the numerator. We will write the result in terms of the scalar integrals using the notation of \cite{Ellis:2007qk}
\begin{align}
\boldsymbol{I} = \left\{ I_3(m^2, s, m^2; 0, m^2, m^2), F_2(s,m^2), I_2(m^2;0,m^2) \right\},\label{eq:ttH-integrals}
\end{align}
where $F_2(s,m^2) = I_2(s;m^2,m^2)- I_2(m^2;0,m^2)$. The result is $A^{(1),4d} = \boldsymbol{c}^{(d_s)} \cdot \boldsymbol{I} A^{(0),4d}$ where the integral coefficients are given by
\begin{align}
\boldsymbol{c}^{(d_s)} = \left\{ - 2 (s-2m^2), (d_s - 4)-\frac{8m^2}{s \beta^2}, d_s \right\}\label{eq:4dcoeffs}
\end{align}
and $\beta^2 = 1-\tfrac{4m^2}{s}$ and $d_s$ is the polarisation state dimension. Using the relation between $\gamma^\mu$ and $\Sigma$- and $\tilde{\Sigma}$-matrices (\ref{eq:gammarelation}) we rewrite the numerator by insertion of $\boldsymbol{1}^A_{\ B} = -\tilde{\Sigma}^{5,AX} \Sigma^5_{XB}$ in (\ref{eq:feynmancalc}):
\begin{align}
N^{4d}  & = \bar{u}_1 \gamma^\mu (\gamma \cdot \bar{\ell}_3 + m)(\gamma \cdot \bar{\ell}_2 + m) \gamma_\mu v_2 \nonumber \\
& = \bar{u}_1 \boldsymbol{1} \gamma^\mu (\gamma \cdot \bar{\ell}_3 + m) \boldsymbol{1} (\gamma \cdot \bar{\ell}_2 + m) \boldsymbol{1} \gamma_\mu v_2 \nonumber \\
& = - \bar{u}_1 \tilde{\Sigma}^5 \Sigma^\mu (\tilde{\Sigma}^\nu \bar{\ell}_{3\nu} - \tilde{\Sigma}^5 m) \Sigma^5 (\tilde{\Sigma}^\rho \bar{\ell}_{2\rho} - \tilde{\Sigma}^5 m) \Sigma_\mu v_2 \nonumber \\
& = \lambda_1 \Sigma^\mu (\tilde{\Sigma} \cdot \ell_3) \Sigma^5 (\tilde{\Sigma} \cdot \ell_2) \Sigma_\mu \lambda_2.
\end{align}
Note the leftover $\Sigma^5$ which is associated with the scalar interaction. Hence the tree level amplitude in six dimensions is given by
\begin{align}
A^{(0),6d} = \lambda_1 \Sigma^5 \lambda_2.
\end{align}
When performing the calculation using six dimensional formalism, the contraction of Lorentz indices from internal gluon lines will include contributions from the extra dimensions. The procedure of reducing the sum over internal states allows us to obtain explicit dependence on space-time dimensionality. This is discussed in section \ref{sec:statesum}. The numerator in the six dimensional calculation is:
\begin{align}
N^{6d}&= \lambda_1 \Sigma^M (\tilde{\Sigma} \cdot \ell_3) \Sigma^5 (\tilde{\Sigma} \cdot \ell_2) \Sigma_M \lambda_2.
\end{align}
The extra contributions appearing from Lorentz contraction in six dimensions are:
\begin{align}
N^{6d}_{\phi_1}&=  - \lambda_1 \Sigma^4 (\tilde{\Sigma} \cdot \ell_3) \Sigma^5 (\tilde{\Sigma} \cdot \ell_2) \Sigma^4 \lambda_2, \\
N^{6d}_{\phi_2}&=  - \lambda_1 \Sigma^5 (\tilde{\Sigma} \cdot \ell_3) \Sigma^5 (\tilde{\Sigma} \cdot \ell_2) \Sigma^5 \lambda_2. \nonumber
\end{align}
It follows from (\ref{eq:6dlagrangian}) that contributions from the scalars can equivalently by obtained with
\begin{align}
N^{6d}_{\phi_1}&=  -\lambda_1 (\Sigma \cdot \ell_3) \tilde{\Sigma}^5 (\Sigma \cdot \ell_2) \lambda_2, \\
N^{6d}_{\phi_2}&= \lambda_1 \gamma_5 (\Sigma \cdot \ell_3) \tilde{\Sigma}^5 (\Sigma \cdot \ell_2) \tilde{\gamma}_5 \lambda_2 \nonumber,
\end{align}
where $(\tilde{\gamma}_5)^A_{\ B} = -i \tilde{\Sigma}^{4,AX} \Sigma^5_{XB}$.
Using the integral basis in (\ref{eq:ttH-integrals}) the result is
\begin{align}
A^{(1),6d}&=  \boldsymbol{c}^{6d} \cdot \boldsymbol{I} A^{(0),6d}\nonumber \\
&= \left\{ -2s , - \frac{16 m^2}{s \beta^2} , 4 \right\} \cdot \boldsymbol{I} A^{(0),6d}, \label{eq:gcontr}\\
A^{(1)}_{\phi_1}&=  \boldsymbol{c}_{\phi_1} \cdot \boldsymbol{I} A^{(0),6d}\nonumber \\
&= \left\{ 0 , 1 , 1 \right\} \cdot \boldsymbol{I} A^{(0),6d}, \label{eq:s4contr} \\
A^{(1)}_{\phi_2}&=  \boldsymbol{c}_{\phi_2} \cdot \boldsymbol{I} A^{(0),6d}\nonumber \\
&= \left\{ -4m^2 , -1 - \frac{8 m^2}{s \beta^2},-1 \right\} \cdot \boldsymbol{I} A^{(0),6d}.\label{eq:s5contr}
\end{align}
The coefficients above are the ingredients needed to perform the state-sum reduction and reproduce (\ref{eq:4dcoeffs}).

\section{Generalised unitarity cuts in six dimensions \label{sec:6Dcuts}}

\begin{figure}[b]
\centering
\[ A^{[L]}\left( 1_{t},2,3,4_{\tb}\right) = \includegraphics[scale=0.2, trim=0 140 0 0]{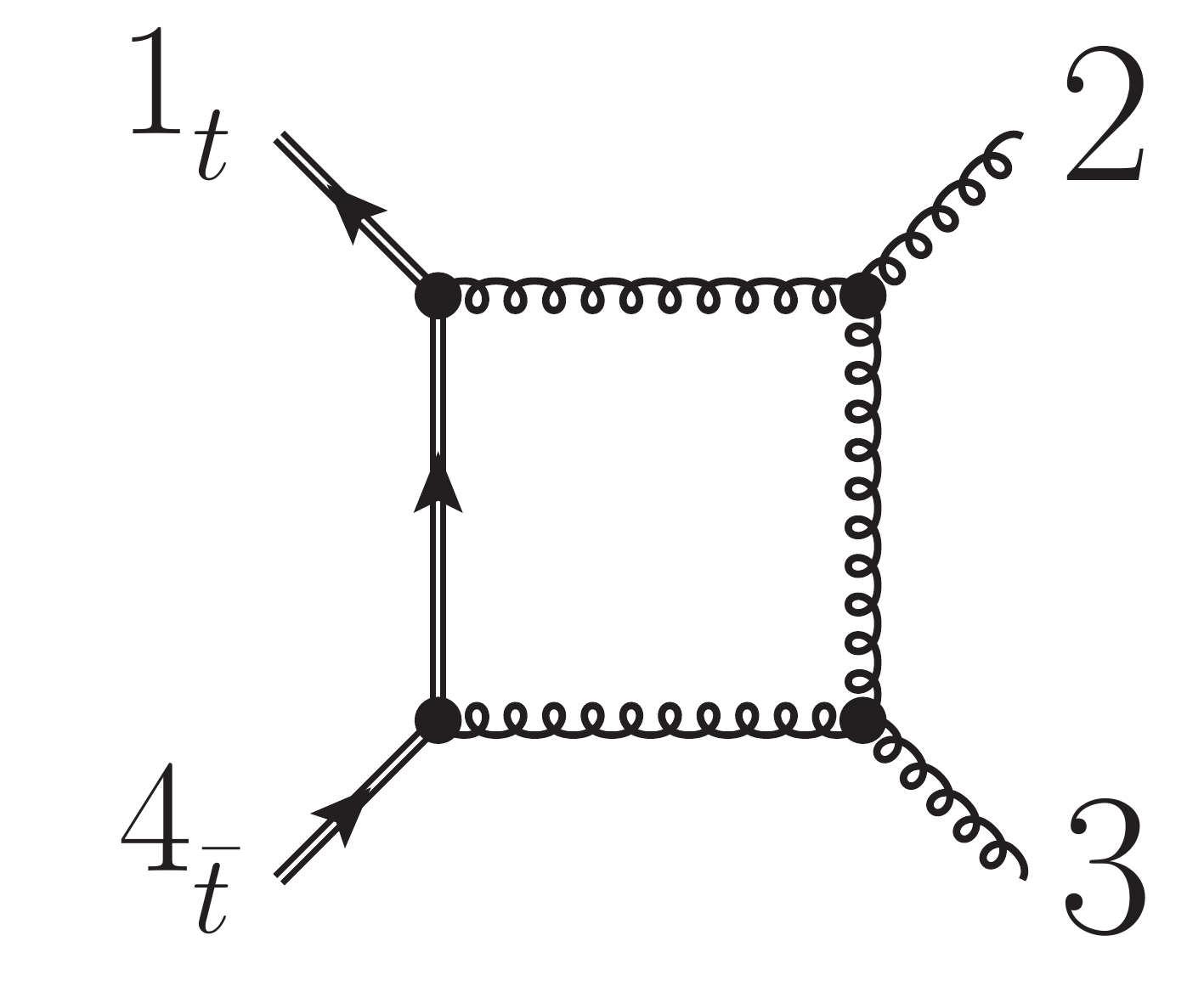} + \dots, \quad \quad
A^{[R]}\left( 1_{t},2,3,4_{\tb}\right) = \includegraphics[scale=0.2, trim=0 140 0 0]{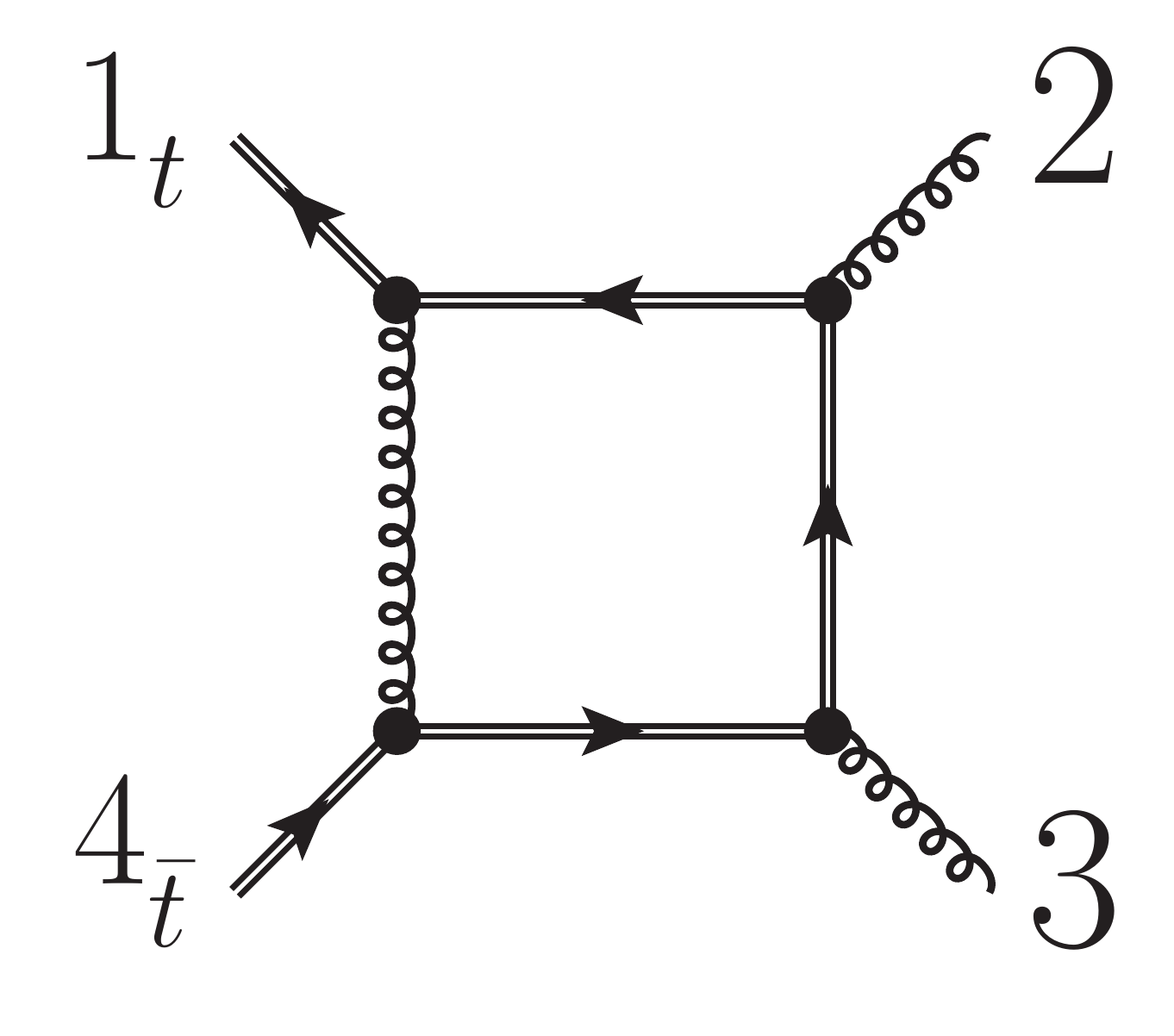} + \dots\] \
\caption{Configurations for left- and right-moving primitive amplitudes contributing to $gg\to t\tb$ scattering.}
\label{fig:LRconfigurations}
\end{figure}

To illustrate our method we consider two gauge invariant primitive amplitudes
relevant for the $gg \to t \tb$ one-loop scattering amplitude. Helicity
amplitudes for this process have been previously presented in
reference~\cite{Badger:2011yu}. Using the usual colour
decomposition~\cite{Bern:1994fz} we define the ordered partial amplitudes
$A^{(1)}_{4;1}$ and $A^{(1)}_{4;3}$ by,
\begin{equation}
\mathcal{A}^{(1)}\left( 1_{t},2,3,4_{\tb}\right) = \sum_{P(2,3)}\left( T^{a_{2}}T^{a_{3}} \right)_{i_{1}}^{\bar{i}_{4}} A^{(1)}_{4;1}\left( 1_{t},2,3,4_{\tb}\right)+
						\tr\left( T^{a_{2}}T^{a_{3}} \right) \delta_{i_{1}}^{\bar{i}_{4}} A^{(1)}_{4;3}\left( 1_{t},4_{\tb};2,3\right),
\end{equation}
where $P(2,3)$ is the permutations over the order of gluons. These partial amplitudes can be further decomposed into gauge invariant primitive amplitudes,
\begin{align}
A^{(1)}_{4;1}\left( 1_{t},2,3,4_{\tb}\right) &= N_c A^{[L]}\left( 1_{t},2,3,4_{\tb}\right) - \frac{1}{N_{c}} A^{[R]}\left( 1_{t},2,3,4_{\tb}\right) \notag \\
					 	 & - N_{f} A^{[f]}\left( 1_{t},2,3,4_{\tb}\right) -  N_{H} A^{[H]}\left( 1_{t},2,3,4_{\tb}\right), \\
 A^{(1)}_{4;3}\left( 1_{t}4_{\tb};2,3\right) &= \sum_{P(2,3)}\left( A^{[L]}\left( 1_{t},2,3,4_{\tb}\right) + A^{[L]}\left( 1_{t},2,4_{\tb},3\right) + A^{[R]}\left( 1_{t},2,3,4_{\tb}\right) \right),
\end{align}
where $N_{c}$ is the number of colours, while $N_{f}$ and $N_{H}$ are the number of
light and heavy fermion flavours, respectively. The left-moving $A^{[L]}$
and right-moving $A^{[R]}$ primitive amplitudes are labelled according to the direction of the fermion current as it enters
the loop, following the convention of reference ~\cite{Bern:1994fz}. Representative diagrams for these
amplitudes are shown in figure \ref{fig:LRconfigurations}. We will not consider the fermion loop contributions $A^{[f]}$ and $A^{[H]}$
in this article as they do not present any further technical difficulties.

\begin{figure}
\centering
\includegraphics[scale=0.3]{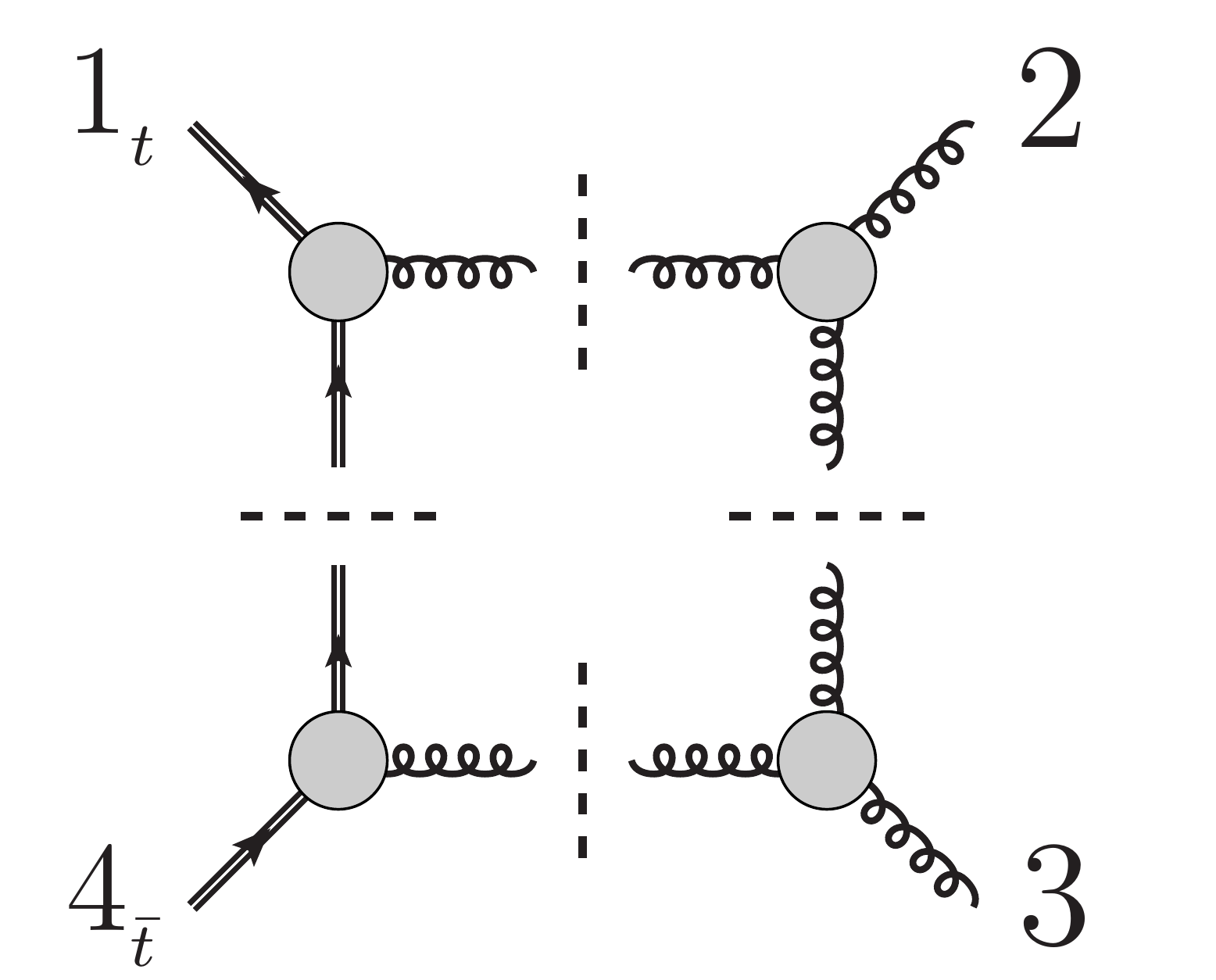} \
\includegraphics[scale=0.3]{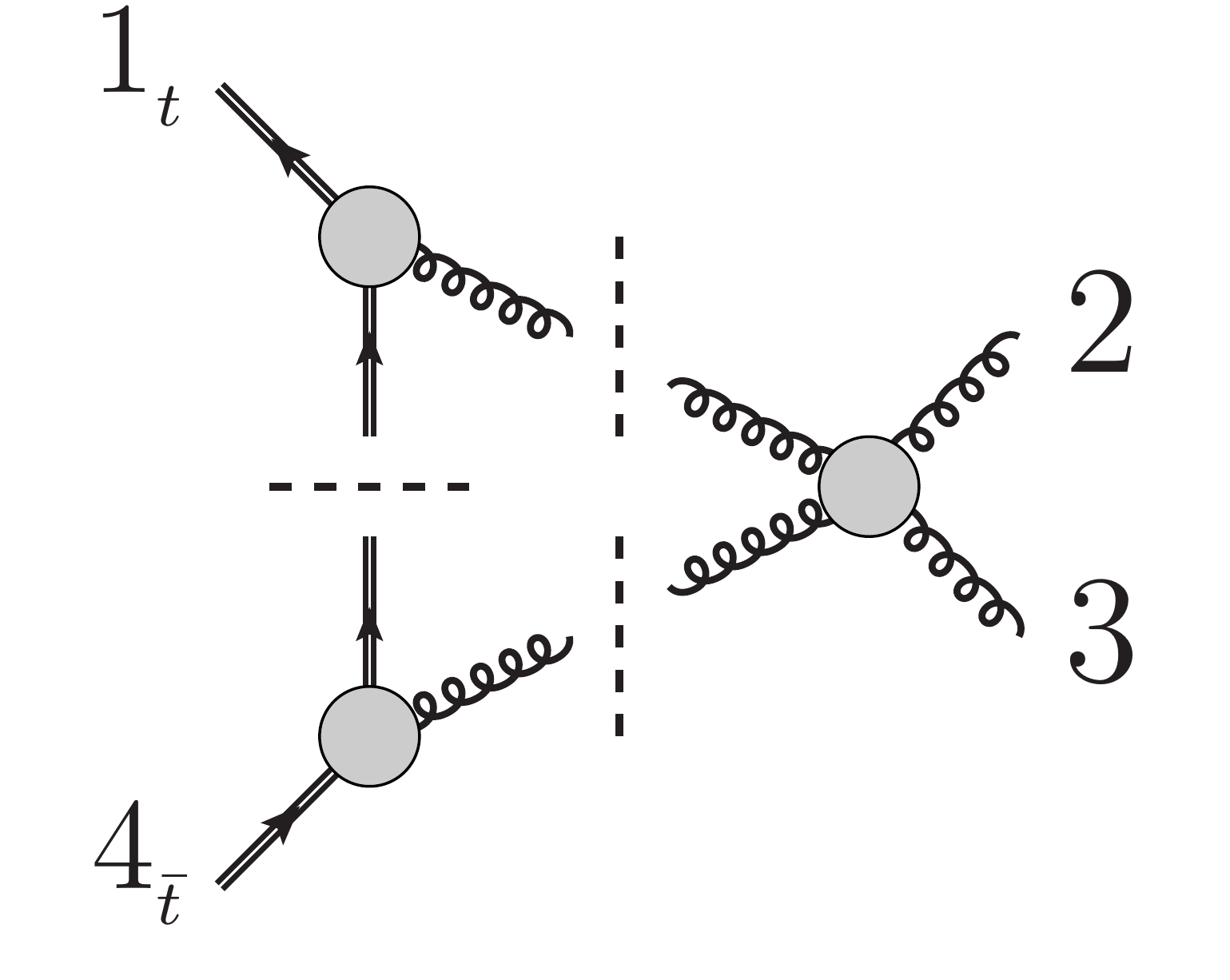} \\
\includegraphics[scale=0.3]{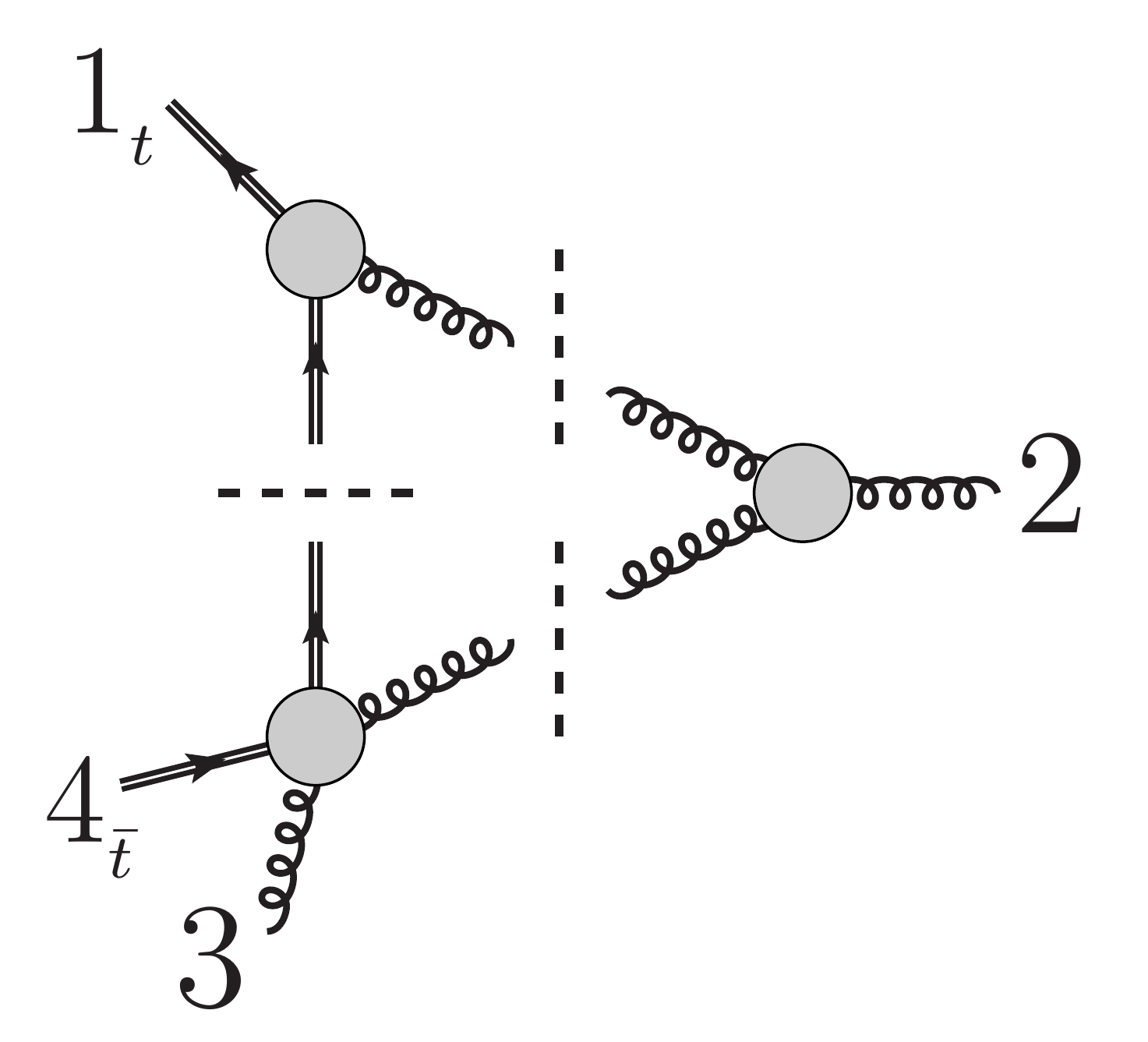} \
\includegraphics[scale=0.3]{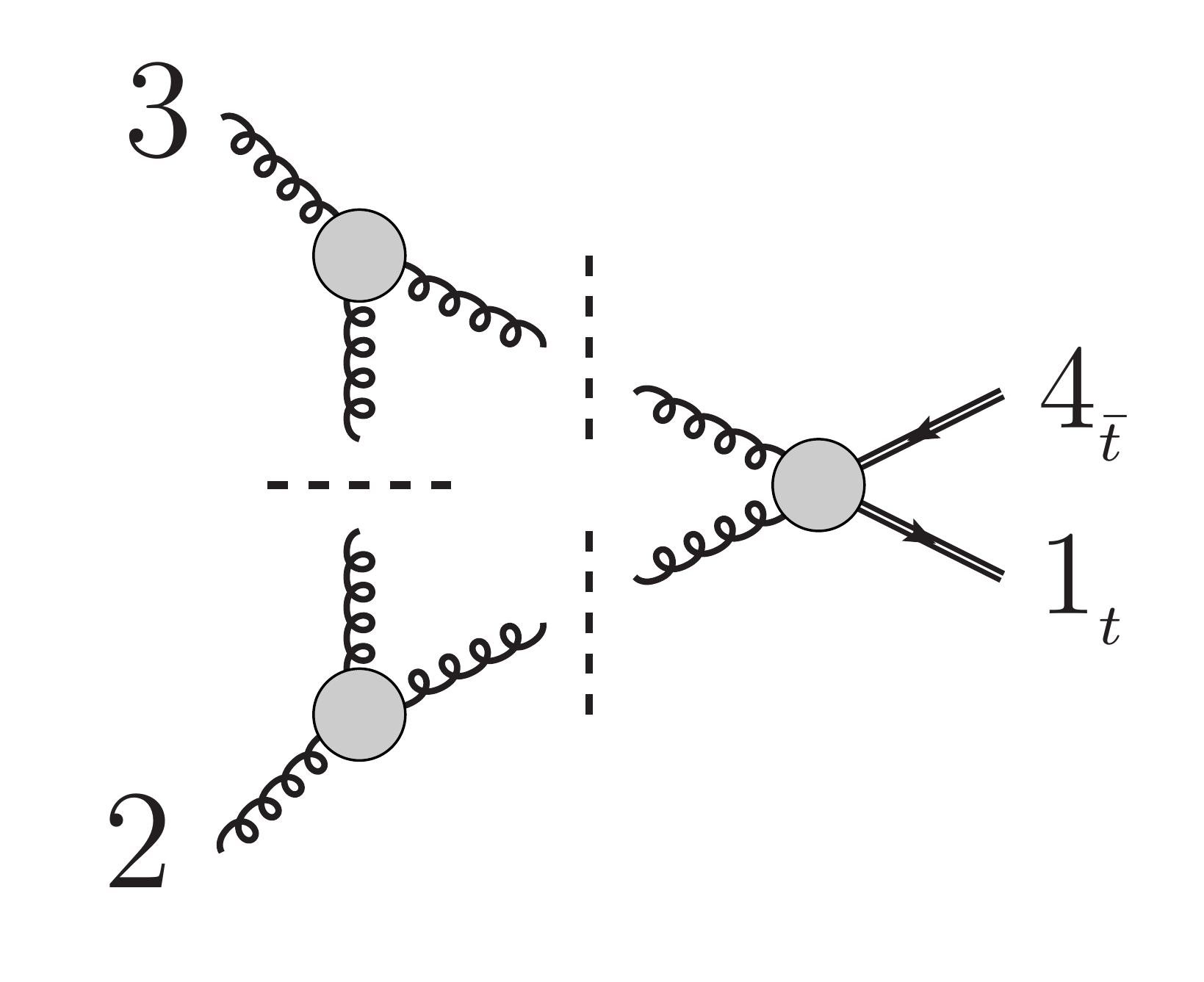} \
\includegraphics[scale=0.3]{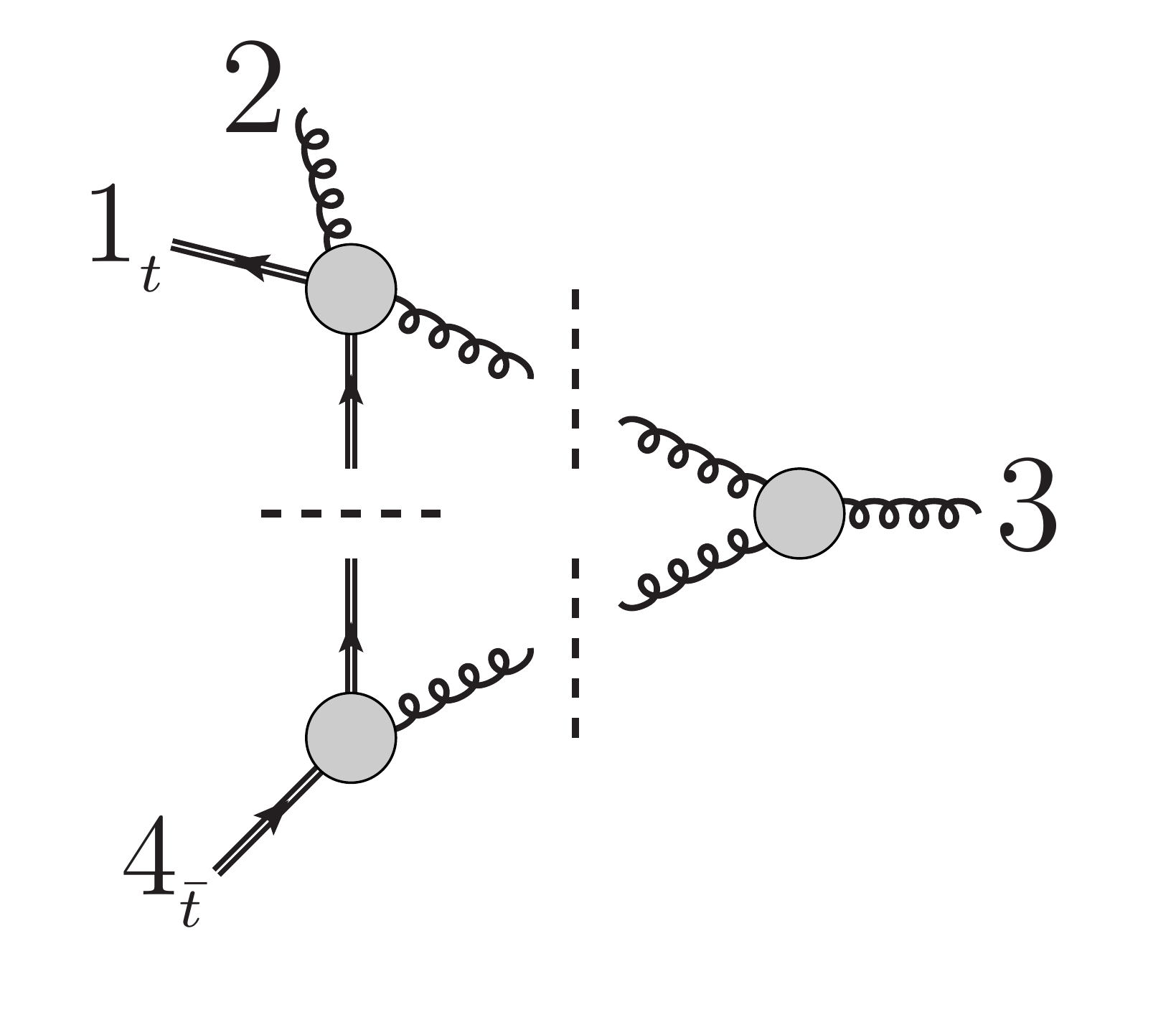} \ \\
\includegraphics[scale=0.3]{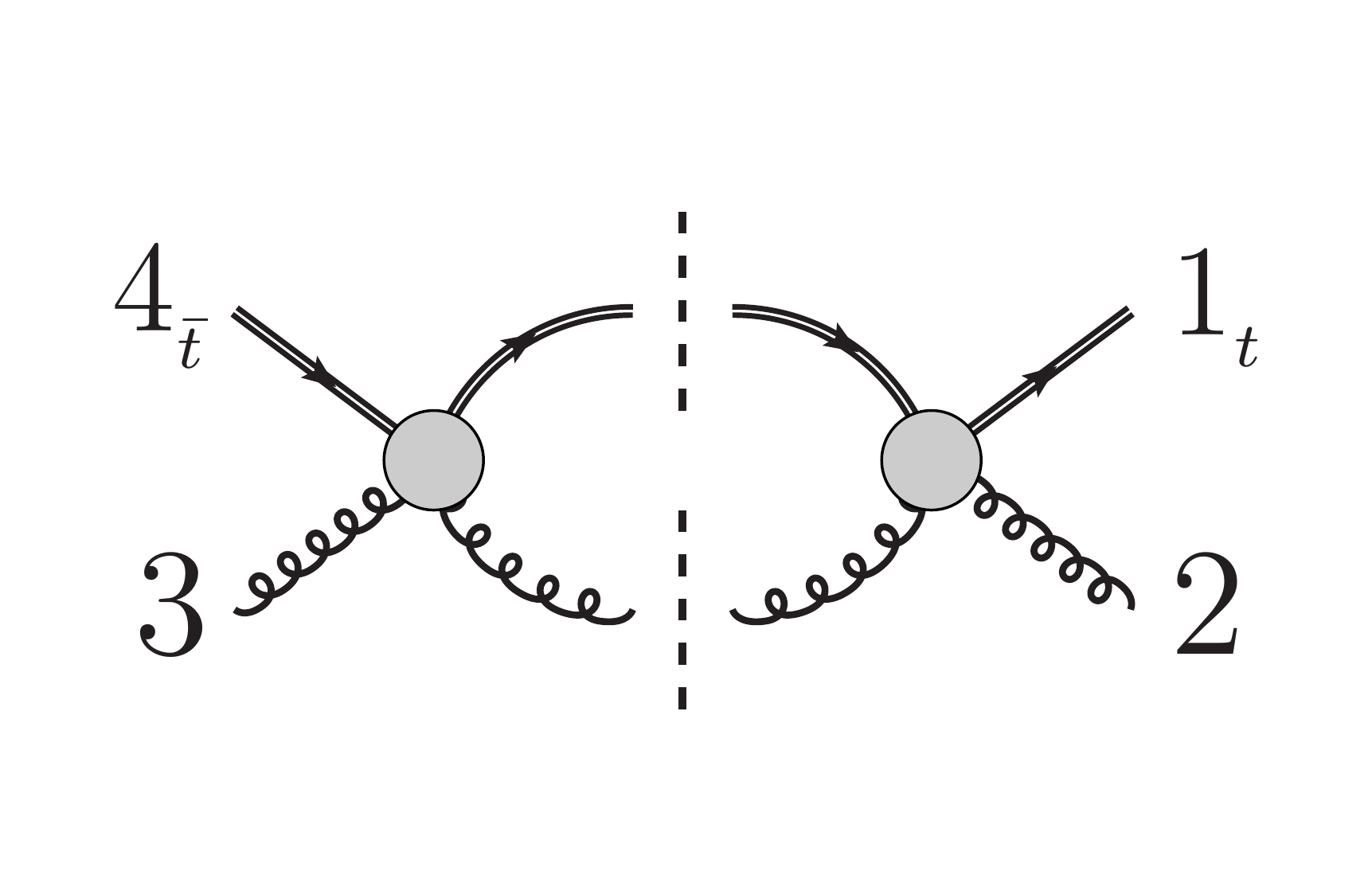} \
\includegraphics[scale=0.3]{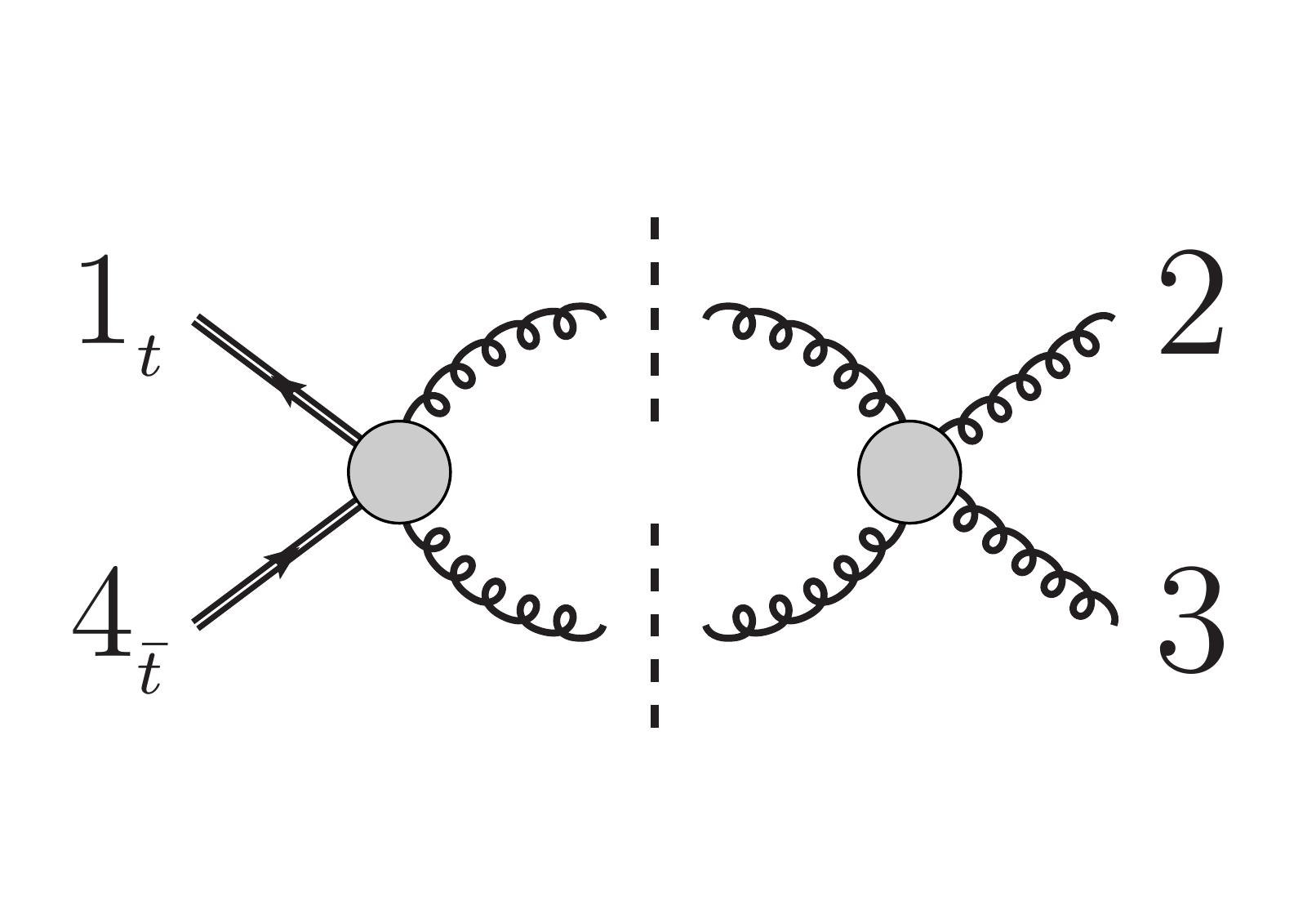}
\caption{The complete set of cuts for $B^{[L]}\left( 1_{t},2,3,4_{\tb}\right)$. Double lines represent massive fermions.}
\label{fig:ALcuts}
\end{figure}

\begin{figure}
\centering
\includegraphics[scale=0.3]{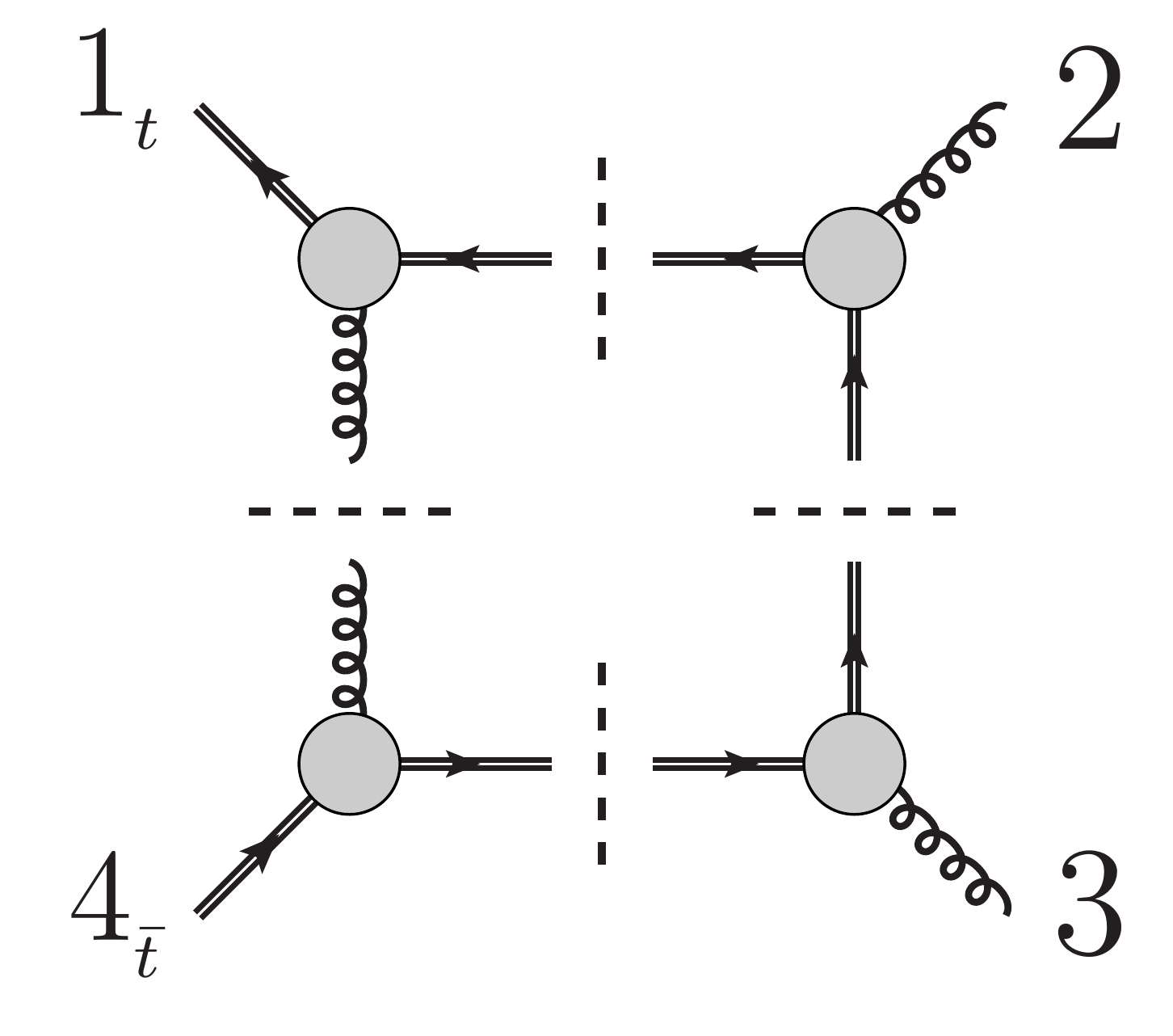} \
\includegraphics[scale=0.3]{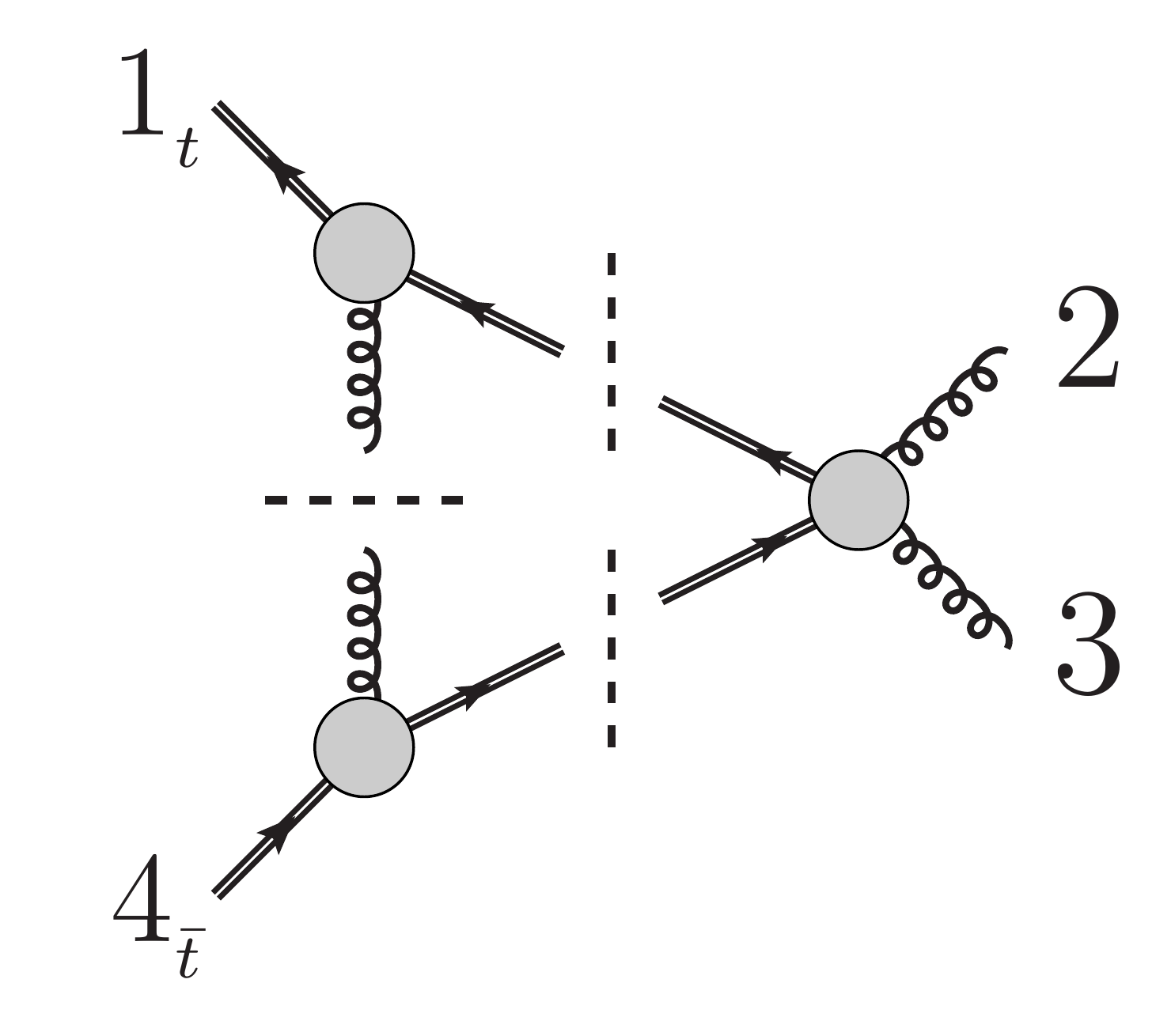} \\
\includegraphics[scale=0.3]{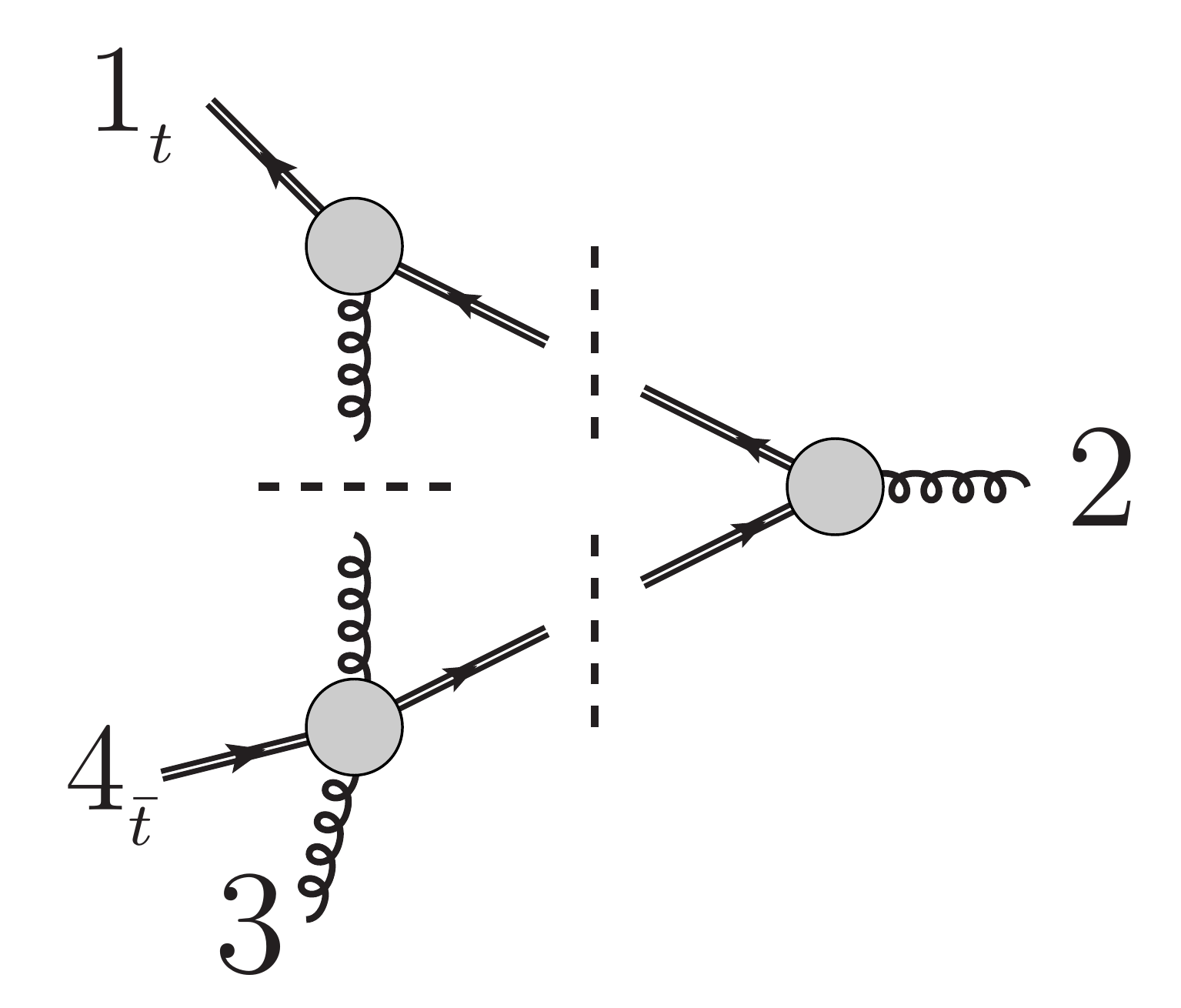} \
\includegraphics[scale=0.3]{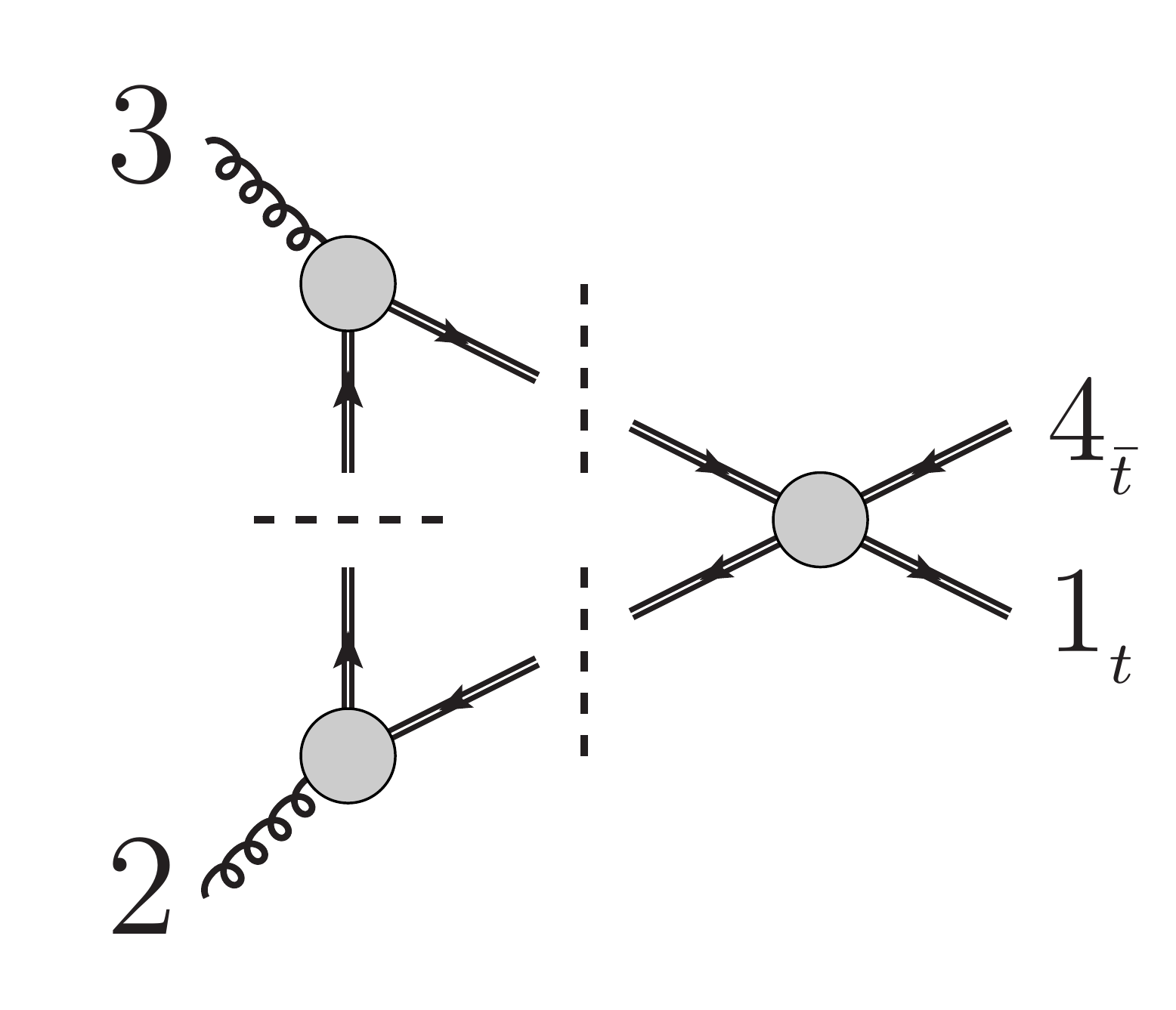} \
\includegraphics[scale=0.3]{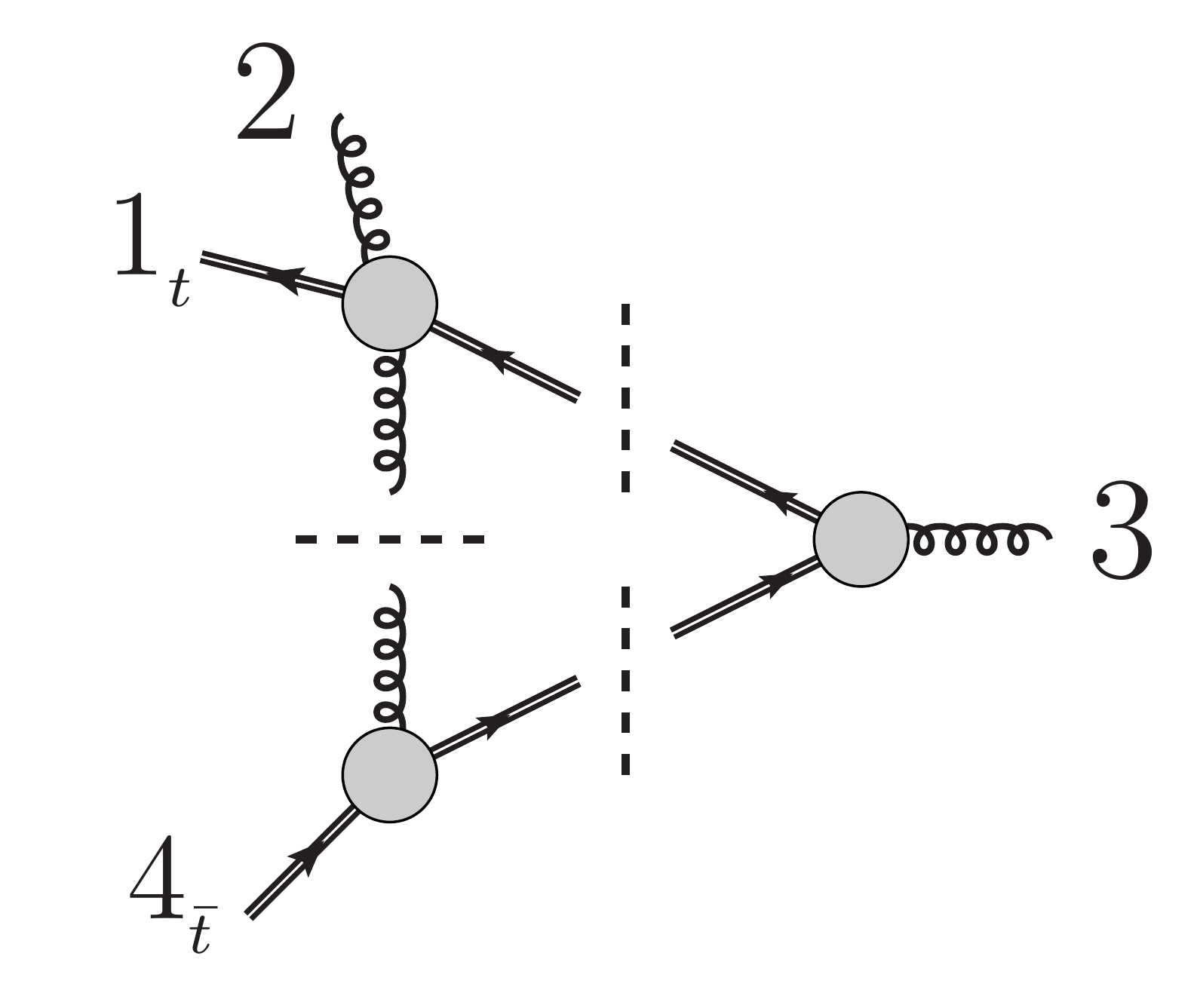} \ \\
\includegraphics[scale=0.3]{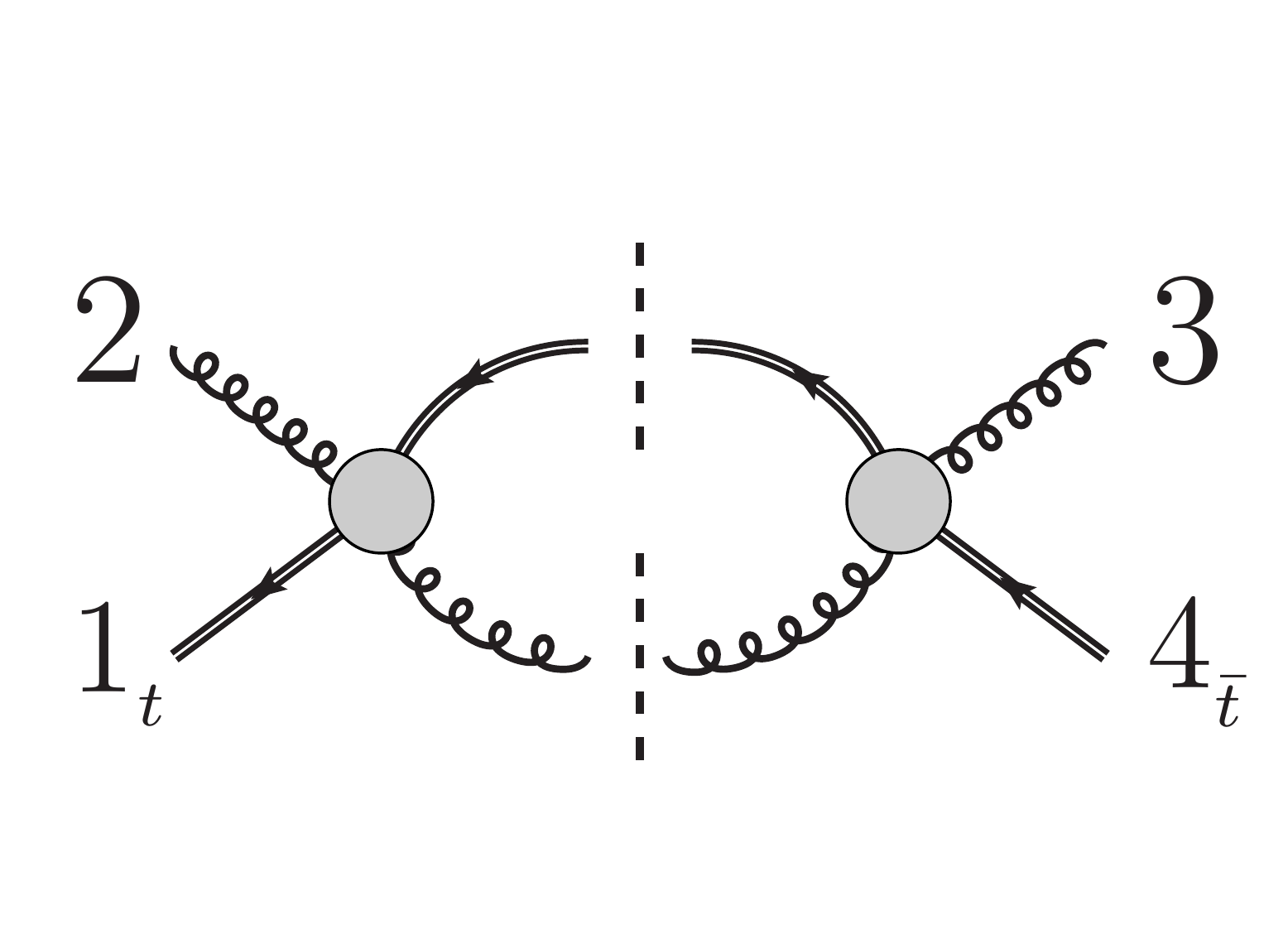} \
\includegraphics[scale=0.3]{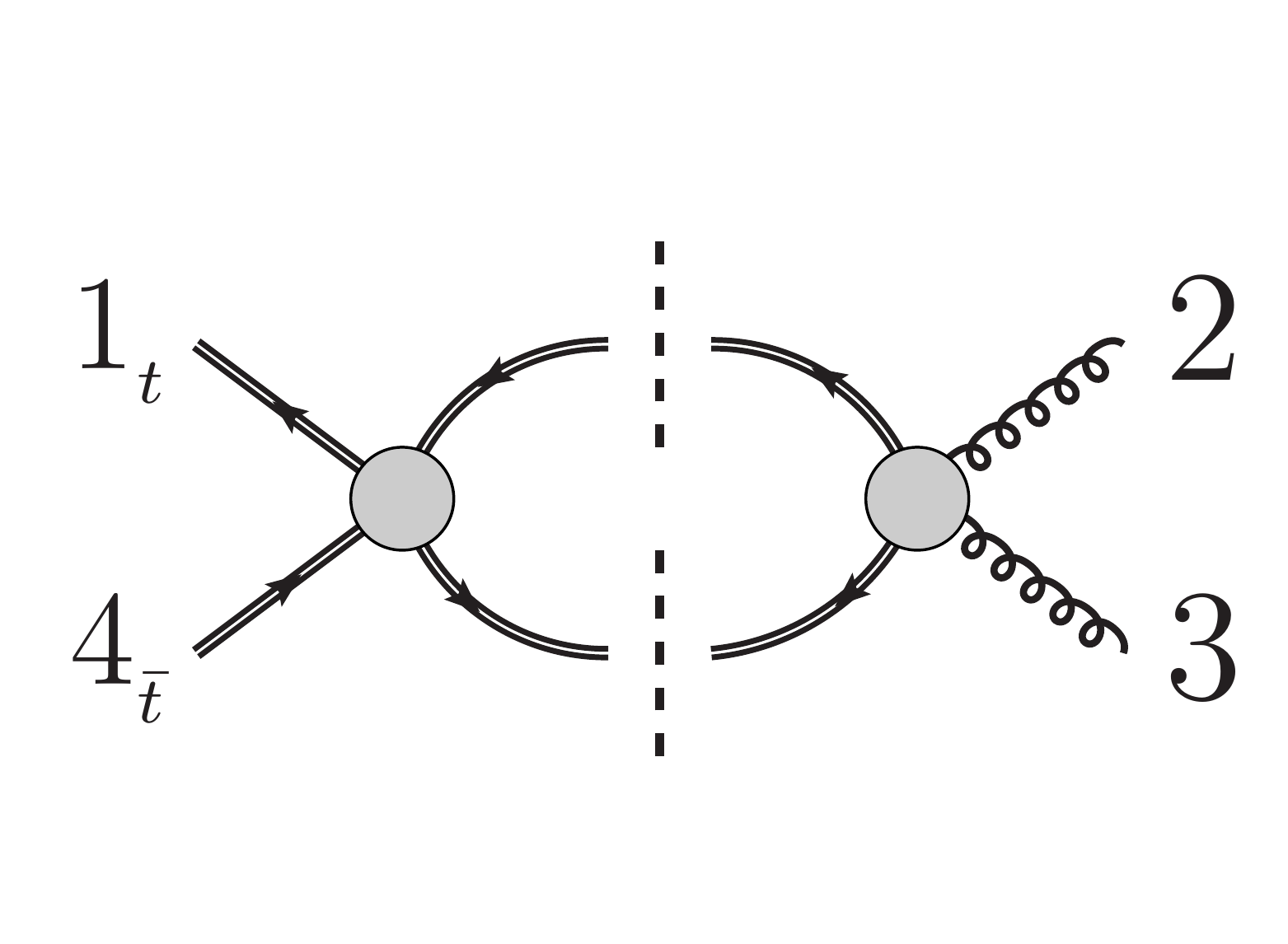}
\caption{The complete set of cuts for $B^{[R]}\left( 1_{t},2,3,4_{\tb}\right)$. Double lines represent massive fermions.}
\label{fig:ARcuts}
\end{figure}

Each primitive amplitude can be decomposed at the integrand level into the basis of integrals described in section
\ref{sec:1lsummary}. To capture the full $d$-dimensional dependence, we first compute generalised cuts in six dimensions
using the spinor-helicity formalism described in the previous section. We then compute the two additional scalar loop
contributions and perform the state sum reduction onto a general dimension $d$ according to eq. \ref{eq:dimred}.
The complete set of generalised cuts needed for the amputated primitives $B^{[L]}$ and $B^{[R]}$, c.f. $\Adcc$ in equation~\ref{eq:1lampdecomp}, are shown in figure \ref{fig:ALcuts}
and \ref{fig:ARcuts}, in which the divergent two-particle and one-particle cuts are removed.

Each six-dimensional cut is associated with a set of loop momenta $\ell_i$ which enter the tree-level
amplitudes. These momenta are determined by solving the system of on-shell equations $\{\ell_{i}^{2}=0, i \in S\}$.
The complete set of loop momenta for our ordered amplitudes are labelled as,
\begin{align}
\ell_{i}^{\mu}&\equiv \ell_{0}^{\mu}-P_{i}^{\mu}, \quad \quad \quad P_{i}^{\mu}=\sum_{n=1}^{i}p_{n}^{\mu}, \nonumber \\
\ell_{0}^{\mu}&\equiv k^{\mu},
\end{align}
where $p_{n}^{\mu}$ are the external momenta and $k$ is the loop integration momentum.

\begin{figure}
\centering
\includegraphics[scale=0.6, trim=0 0 0 0]{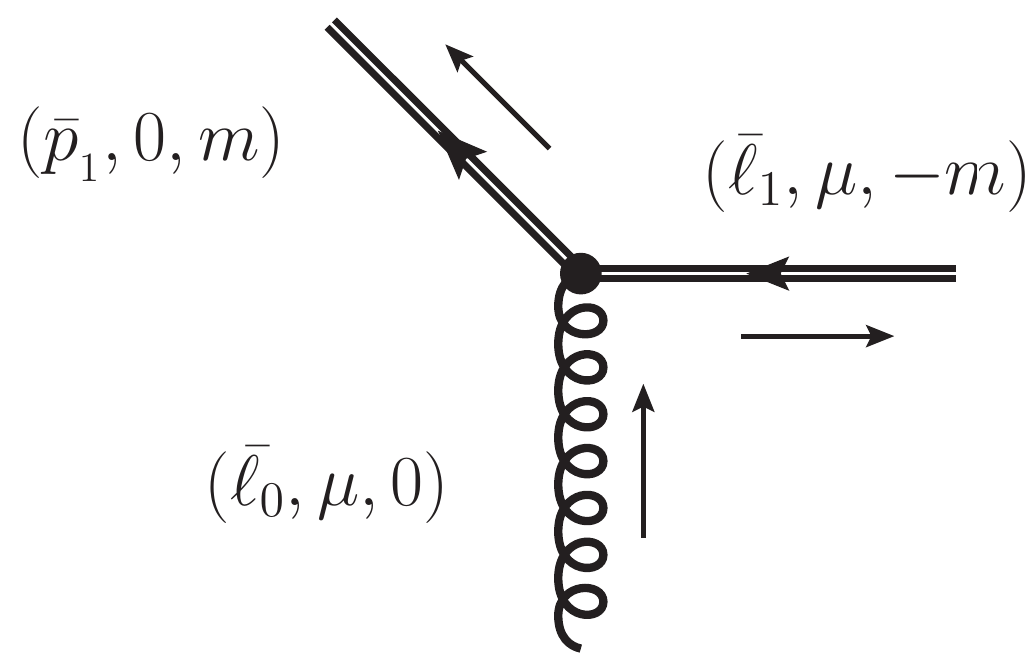}
\caption{To perform the unitarity cuts of the six dimensional propagators
  involving internal fermions, we allow  the $(d-4)$ part, $\mu$, of the loop
  momentum to flow in the fifth component and the mass term to flow in the
  sixth component, in order to easily impose momentum conservation.}
\label{fig:vertex}
\end{figure}

The internal particles are embedded into six dimensions by allowing the mass
to flow in the sixth component, following our convention in eq. \eqref{eq:6dfor4dmass},
and the $(d-4)$ part of the loop momentum to flow in the fifth component,
\begin{align}
\text{gluon loop momentum:}     & \quad \ell = \{\bar{\ell},\mu, 0\}, \nonumber \\
\text{fermion loop momentum:} & \quad \ell = \{\bar{\ell}, \mu, m \}.
\label{eq:loopmoms}
\end{align}
The gluon and fermion loop propagators can then be expanded into a four-dimensional
part and an effective mass term $\mu^2$,
\begin{align}
\text{gluon propagator:}     & \quad \ell^{2} = \bar{\ell}^{2} - \mu^{2},\\
\text{fermion propagator:} & \quad \ell^{2} = \bar{\ell}^{2} - \mu^{2} - m^{2}.
\label{eq:props}
\end{align}
This choice is particularly convenient when requiring momentum conservation and
orthogonality of the $-2\eps$  component with respect to the external massive
fermions momenta expressed in the six dimensional representation, as shown in
figure \ref{fig:vertex}.

As an explicit example we will describe the computation of the quadruple cuts.
The on-shell equations for these cuts in the left- and right-moving configurations are,
\begin{align}
S_{4;0123}^{L} =
\begin{cases}
\ell_{0}^{2}=\ell_{1}^{2}=\ell_{2}^{2}=\ell_{3}^{2}=0 \\
\ell_{0}^{(5)}=m
\end{cases}, &
& S_{4;0123}^{R} =
\begin{cases}
\ell_{0}^{2}=\ell_{1}^{2}=\ell_{2}^{2}=\ell_{3}^{2}=0 \\
\ell_{0}^{(5)}=0
\end{cases}.
\label{eq:4cutsyst}
\end{align}
The constraint on the sixth component of the loop momentum $\ell_0$ distinguishes
between the two different configurations.

We construct explicit solutions for the six-dimensional spinors
of $\ell_i$ by introducing arbitrary two-component reference spinors $x_a$ and $\tilde{x}_{\dot{a}}$.
These solutions, which have a similar form to those presented in refs.~\cite{Risager:2008yz,Berger:2008sj},
take a simple form,
\begin{align}
  \ell_0^M &=
  \frac{\la x.4 | \Sigma^M \, 1 \, 2 \, 3 | 4.\tilde{x} ]}
  {\la x.4 | 2 \, 3 | 4.\tilde{x}]},
  &
  \ell_1^M &=
  \frac{\la x.4 | 1 \, \tilde{\Sigma}^M \, 2 \, 3 | 4.\tilde{x} ]}
 {\la x.4 | 2 \, 3 | 4.\tilde{x}]},
  \nonumber\\
  \ell_2^M &=
  \frac{\la x.4 | 1 \, 2 \, \Sigma^M \, 3 | 4.\tilde{x} ]}
  {\la x.4 | 2 \, 3 | 4.\tilde{x}]},
  &
  \ell_3^M &=
  \frac{\la x.4 | 1 \, 2 \, 3 \, \tilde{\Sigma}^M | 4.\tilde{x} ]}
  {\la x.4 | 2 \, 3 | 4.\tilde{x}]},
  \label{eq:quadcutsol}
\end{align}
where $\la x.4 | = x^a \la 4_a|$, $| 4.\tilde{x} ] = | 4^{\dot{a}}] \tilde{x}_{\dot{a}}$ and the spinor product strings have the following expression (for $n$ even)
\begin{align}
\la 1_{a} | 2 \, 3\dots \, (n-1) | n^{\dot b} ] = \lambda^{A}_{a}(p_{1}) (\Sigma\cdot p_{2})_{AB} \, (\tilde{\Sigma}\cdot p_{3})^{BC} \dots (\tilde{\Sigma}\cdot p_{n-1})^{CA}\tilde{\lambda}_{A}{}^{ \dot b}(p_{n}).
\end{align}
The expressions for the two reference spinors can generically be chosen to be
\begin{align}
x_{a} = (1, \tau_{1}), & &
\tilde{x}_{\dot{a}} = (1, y),
\end{align}
where $y$ is fixed, for left and right, by the mass constraint for $\ell_{0}^{(5)}$ specified in \eqref{eq:4cutsyst}.
Because we have a system of 5 equations for 6 dimensional momenta, the parameter $\tau_{1}$ is left unconstrained.

On the quadruple cut the amplitudes factorise into products of four tree-level amplitudes,
\begin{equation}
\includegraphics[scale=0.17, trim=70 160 0 0]{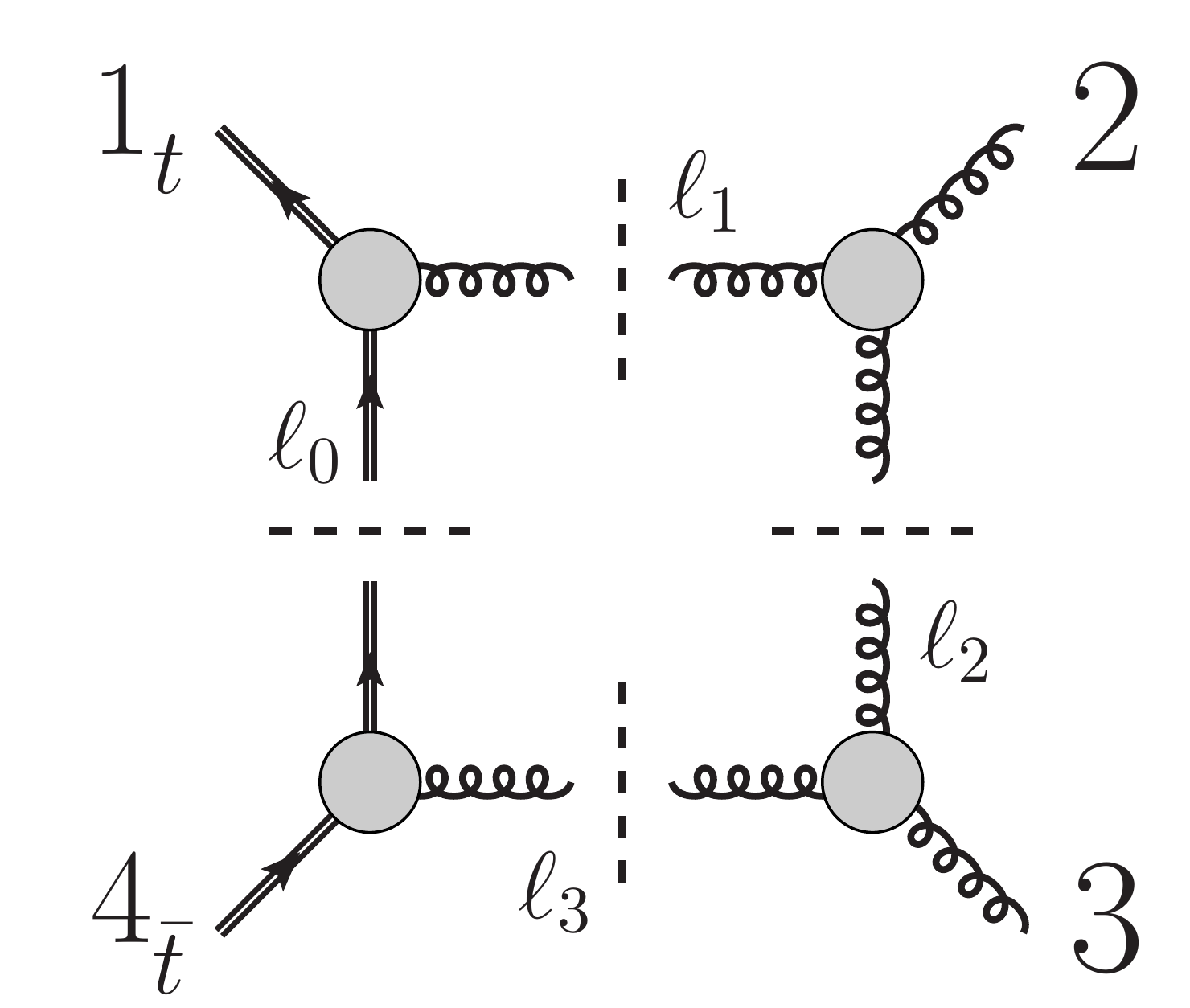} 
C_{4;0123}^{L} = A(-\ell_{0a},1_{t}^{\alpha},\ell_{1}^{b\dot{b}} )A(-\ell_{1b\dot{b}},2^{\beta\dot{\beta}},\ell_{2}^{c\dot{c}} )
A(-\ell_{2c\dot{c}},3^{\gamma\dot{\gamma}},\ell_{3}^{d\dot{d}})A(-\ell_{3d\dot{d}},4_{\tb}^{\delta},\ell_{0}^{a} ),
\\[2em]  
\label{eq:4cutL}
\end{equation}
and
\begin{equation}
\includegraphics[scale=0.17, trim=70 160 0 0]{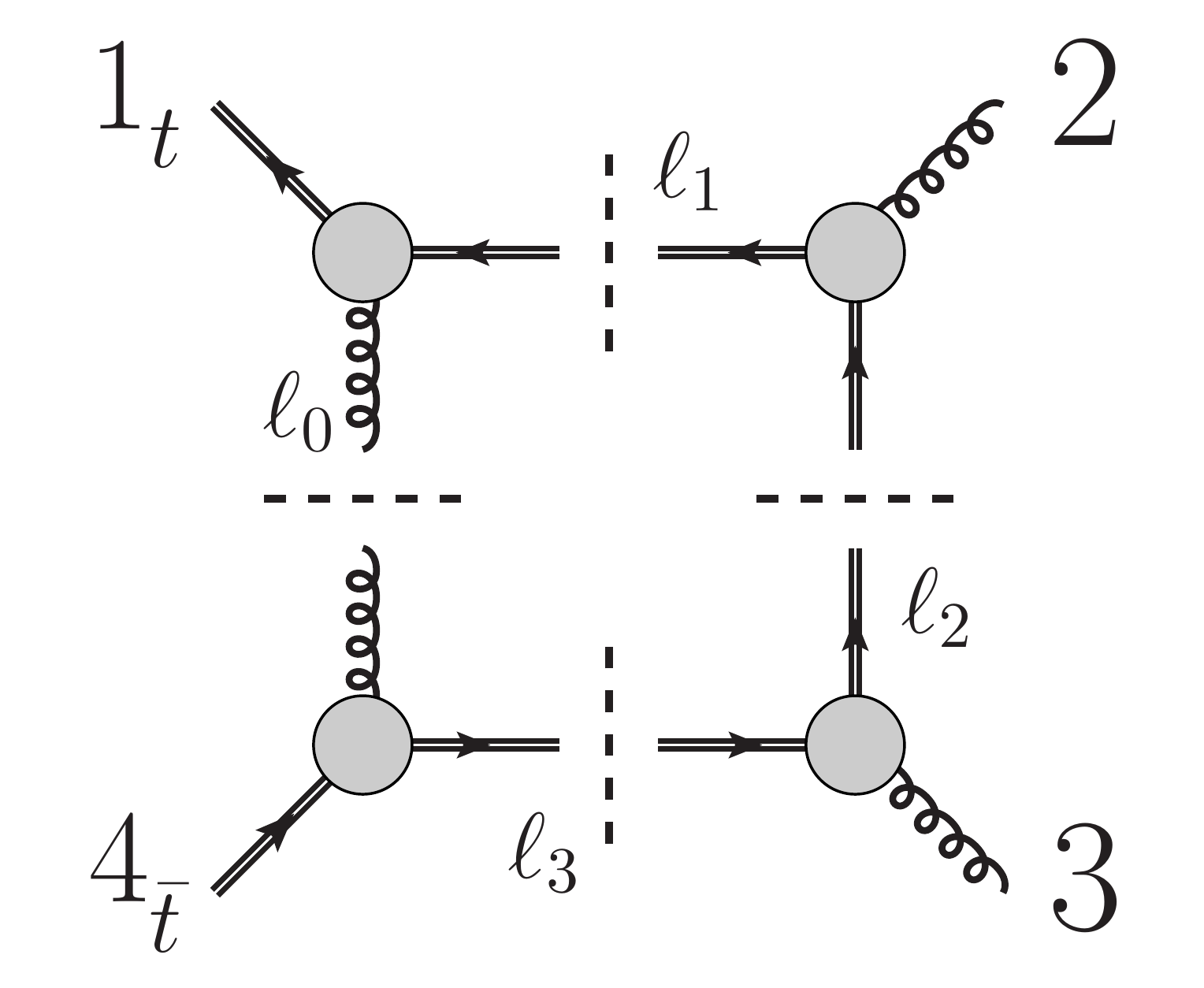}  
C_{4;0123}^{R} = A(-\ell_{0a\dot{a}},1_{t}^{\alpha},\ell_{1}^{b} )A(-\ell_{1b},2^{\beta\dot{\beta}},\ell_{2}^{c} )
A(-\ell_{2c},3^{\gamma\dot{\gamma}},\ell_{3}^{d})A(-\ell_{3d},4_{\tb}^{\delta},\ell_{0}^{a\dot{a}} ),
\\[2em]
\label{eq:4cutR}
\end{equation}
where in both cases the repeated $SU(2)$ spinor indices are summed over the six dimensional polarisation states.

The integrand reduction method then proceeds to extract the five independent
coefficients in the integrand parametrisation from eq.
\eqref{eq:6dintegralbasis} by evaluating both the product of trees and the
irreducible scalar products $\mu^2$ and $k\cdot w_{1;123}$ using the on-shell
solution in eq. \eqref{eq:quadcutsol} and comparing the resulting rational
functions in $\tau_1$. We encounter an interesting subtlety when following this procedure since the
six-dimensional cut contains additional terms which are linear in the
extra-dimensional component of the loop momentum $\mu$. These terms are
spurious and integrate to zero, but require additional coefficients to be added at the integrand level
if this direct approach is taken. A slightly simpler approach is to
cancel the linear part of the cut by averaging over the two different flows of the momentum in the
fifth component,
\begin{align}
  \frac{1}{2} \left( C_{4;0123}\big|_{S^{+}_{4;0123}} + C_{4;0123}\big|_{S^{-}_{4;0123}} \right)
  & =
  \Delta_{\{0,1,2,3\}}\big|_{S_{4;0123}} \label{eq:intredbox},
\end{align}
where
\begin{align}
   S^{+} &= \left\{ \ell_{i}^{2}=0,  \ell_{i} = \{\dots,\mu,\dots\} \right\},
&  S^{-} &= \left\{ \ell_{i}^{2}=0,  \ell_{i} =  \{\dots,-\mu,\dots\} \right\}.
\end{align}

The triangle and bubble coefficients follow using the OPP method to systematically
remove all singularities from the cut amplitude using the previously computed
irreducible numerators. The mass dependence of the propagators is now dictated
by six dimensional momentum conservation applied to the loop momenta, so all
propagators are simply $\ell_i^2$. To remove the terms linear in $\mu$, 
we average over the two directions for the extra-dimensional component, as described above.
Thus,
\begin{subequations}
\begin{align}
  \frac{1}{2} \sum_{\sigma=\pm} C_{4;0123} \big|_{S^\sigma_{4;0123}} & = \Delta_{\{0,1,2,3\}}\big|_{S_{4;0123}},\\
  \frac{1}{2} \sum_{\sigma=\pm} C_{3;012} \bigg|_{S^\sigma_{3;012}}
   - \frac{ \Delta_{\{0,1,2,3\}}}{ \ell_{3}^{2} }
   \bigg|_{S_{3;012}}
   & =
   \Delta_{\{0,1,2\}} \big|_{S_{3;012}},   \\
  \frac{1}{2} \sum_{\sigma=\pm} C_{2;02} \bigg|_{S^\sigma_{3;02}}
   - \left(
     \frac{\Delta_{\{0,1,2\}} }{ \ell_{1}^{2} }
   + \frac{\Delta_{\{0,2,3\}} }{ \ell_{3}^{2} }
   + \frac{\Delta_{\{0,1,2,3\}}}{ \ell_{1}^{2}\ell_{3}^{2} }
   \right) \bigg|_{S_{2;02}}
   & =
   \Delta_{\{0,2\}} \big|_{S_{2;02}},
\end{align}
\label{eq:coefficients}%
\end{subequations}
where the parametrisations for each irreducible numerator are those of equation
\eqref{eq:deltaparam}. The remaining triple and double cuts follow by
permuting the equations \eqref{eq:coefficients}. Further details on the on-shell cut solutions
are given in appendix \ref{app:cutsols} and a full set of numerical results for
the six dimensional cuts are listed in the \textsc{Mathematica} notebook
accompanying this article.

The final step to dimensionally reduce the coefficients from 6 to a general
dimension $d$ is to remove the extra degrees of freedom contained in the six
dimensional loop momentum according to eq. \eqref{eq:dimred}. The computation
of these extra cuts is done using the same procedure as above, but the internal
gluon lines in figures~\ref{fig:ALcuts} and \ref{fig:ARcuts} are replaced with
scalar lines. For example, the quadruple cuts are given by the following expressions
\begin{align}
C_{4;0123}^{L, \phi_{(1,2)}} & = A(-\ell_{0\dot{a}},1_{t}^{\alpha},\ell_{1} )A(-\ell_{1},2^{\beta\dot{\beta}},\ell_{2} )
A(-\ell_{2},3^{\gamma\dot{\gamma}},\ell_{3})A(-\ell_{3},4_{\tb}^{\delta},\ell_{0}^{\dot a} ), \\
C_{4;0123}^{R, \phi_{(1,2)}} &= A(-\ell_{0},1_{t}^{\alpha},\ell_{1}^{\dot b} )A(-\ell_{1\dot b},2^{\beta\dot{\beta}},\ell_{2}^{\dot c} )
A(-\ell_{2\dot{c}},3^{\gamma\dot{\gamma}},\ell_{3}^{\dot d})A(-\ell_{3 \dot d},4_{\tb}^{\delta},\ell_{0} ).
\end{align}
A complete set of fermion and scalar integrand coefficients are presented in the attached notebook.

\section{Determining the remaining integral coefficients \label{sec:wftadcoeff}}

At this point, let us pause to take stock of what has been achieved, and what remains to be done. To do so, we return to equation~\eqref{eq:1lampdecomp}, the standard expression for a one-loop amplitude, expanded in a basis of scalar integrals:
\begin{equation}
A_n^{(1)} = \Adcc + c_{2;m^2} I_{2,m^2} + c_1 I_1.
\label{eq:1lampdecompRepeat}
\end{equation}
By definition, $\Adcc$ is the part of the amplitude which can be computed using finite $d$-dimensional unitarity cuts; its expansion in terms of an integral basis was explicitly given in equation~\eqref{eq:6dintegrandbasis}. We have therefore computed $\Adcc$ explicitly in section~\ref{sec:6Dcuts}. A complete construction of the amplitude requires us to find the integral coefficients $c_{2;m^2}$ and $c_1$. This is the task of the present section. 

\subsection{Fixing $c_{2,m^2}$ by matching the poles in $4-2\eps$ dimensions \label{sec:wfbub}}

Our first source of additional information is the universal pole structure of four dimensional amplitudes.
The poles of general one-loop QCD amplitudes in four dimensions were inferred from the corresponding
real-radiation contributions to the NLO cross-section in full generality by
Catani, Dittmaier and Trocsanyi \cite{Catani:2000ef},
\begin{align}
  A^{(1),4-2\eps} = c_\Gamma \, I^{(1)}(\eps) A^{(0)} + \text{finite}.
  \label{eq:universal4dPoles}
\end{align}
The integrals $I_{2,m^2}$ and $I_1$ appearing in equation~\eqref{eq:1lampdecompRepeat} are divergent, and therefore the coefficients $c_{2;m^2}$ and $c_1$ contribute to the pole structure of our amplitude. This will allow us to constrain them.

For the simplified case of $t\tb+n(g)$ with $n_f$ light quark flavours and one heavy flavour of
mass $m$, the function $I^{(1)}(\eps)$ appearing in the univeral pole formula is, explicitly,
\begin{align}
  I^{(1)}(\eps) = \frac{ n_g \beta_0(n_f+1)}{2\eps}
  + \sum_{i,j} \left( \frac{\mu_R^2}{s_{ij}} \right)^\eps \mathcal{V}_{ij}
  - n_g \Gamma_g - 2 \Gamma_t + \text{finite}.
 \label{eq:DCT}
\end{align}
Following Catani et al.~\cite{Catani:2000ef}, this formula corresponds to partially renormalised
amplitudes. The first term contains UV poles related to charge renormalisation, the second term
corresponds to soft-collinear poles and takes the familiar dipole form in colour space. The last terms
contain poles given by the anomalous dimensions,
\begin{align}
  \Gamma_g &= \frac{\beta_0(n_f)}{2 \eps} + \frac{2T_R}{3} \log\left( \frac{\mu_R^2}{m_t^2} \right),\\
  \Gamma_t &= C_F\left( \frac{1}{\eps} - \frac{1}{2} \log\left(\frac{\mu_R^2}{m_t^2} \right) -2 \right).
\end{align}
The QCD $\beta$ function appears as a function of the active fermion flavours $\beta_0(n_f) = (11
C_A - 4 T_R n_f)/3$. For the purposes of this paper we will not require the finite parts of
$I^{(1)}$ which depend on the dimensional regularisation scheme, e.g. CDR or FDH/DR.  The exact form
of the function $\mathcal{V}$ is a little more complicated and not of direct relevance here. Clearly
there is an enormous amount of information contained in this result and further details can be found
by consulting the original reference \cite{Catani:2000ef}.

The simple observation relevant for our approach is that this universal information
can be compared to the integral basis in equation \eqref{eq:1lampdecompRepeat}, enabling a partial determination of
the unknown coefficients of wavefunction bubble and tadpole integrals. These integrals
give rise to single poles in $\eps$ and single logarithms in the mass $m$. This comparison
is however insufficient to constrain both $c_{2,m^2}$ and $c_1$.

It is convenient to modify the integral basis slightly, introducing finite bubble and tadpole functions defined by
\begin{align}
  F_{2;i_1,i_2} &= I_{2,i_1,i_2} - I_{2,m^2}, \\
  F_1 &= I_1 - m^2 I_{2,m^2}.
\end{align}
The result of this modification is that only the finite bubble integrals and the wavefunction integral contribute to the $\log(\mu_{R}^{2}/m_{t}^{2})$ dependence
of the universal pole structure \eqref{eq:DCT}.
Upon matching the amplitude with the universal pole structure, we find that the amplitude takes the explicit expression
\begin{align}
  A^{(1)} = A^{6D,(1)}\Bigg|_{I_2\to F_2} + \frac{d_{s}-2}{4} A^{(0)} I_{2,m^2} + c_1 F_1,
  \label{eq:1l6dbasisshift}
\end{align}
where the only missing information now lies in the tadpole coefficient $c_1$.

\subsection{Counterterms for QCD in six dimensions  \label{sec:qcd6d}}

Because of our exploitation of the universal four-dimensional pole structure, the one-loop amplitude, in the form given in equation~\eqref{eq:1l6dbasisshift}, has the property that its infrared and ultraviolet poles have been correctly determined. In addition, all logs in the mass $m_t$ are correctly reproduced. Indeed, the unknown coefficient $c_1$ now multiplies an integral $F_1$ which we may explicitly compute:
\begin{equation}
F_1  \overset{d=4-2\eps}{=} -i c_\Gamma m^2 + \mathcal{O}(\eps) = -\frac{i m^2}{(4\pi)^2} + \mathcal{O}(\eps).
\end{equation}
Since $c_1$ is also a rational function, the part of the amplitude which remains to be determined is simply a rational function of the external momenta and masses. 

Having made heavy use of higher dimensional methods so far in our computation, it is natural to regard the four-dimensional result we wish to determine as a specialisation of an amplitude that exists in higher dimensions. Indeed, a quantum field theory which is an analogue of QCD exists in six dimensions. Moreover, in six dimensions the integral $F_1$ is no longer simply a finite rational function. It has an epsilon-pole given by
\begin{equation}
F_1  \overset{d=6-2\eps}{=} -\frac{i m^4}{(4\pi)^3} \frac{1}{6\epsilon} + \mathcal{O}(\eps).
\end{equation}
We may therefore find $c_1$ by comparison with the universal epsilon-pole structure of the amplitude in six dimensions.

Thus, we are motivated to consider QCD in six dimensions. Above four dimensions QCD ceases to be renormalisable, so to determine the universal epsilon-pole structure in six dimensions we must include higher (mass-)dimension operators\footnote{It is linguistically unfortunate that we are now dealing with operators of mass-dimension five and six (using the usual four-dimensional classification of operator dimension) in a theory defined in six spacetime dimensions. We hope that context will make the meaning of the word ``dimension'' clear.} and treat the theory as an effective theory. 
By power counting, these operators have one or two powers of momentum more than in the usual QCD Lagrangian, so that they have mass-dimension five or six. The point of view we adopt is that the role of the additional operators is simply to provide counterterms, subtracting the infinities from any one-loop amplitude in the theory. Once all the counterterms have been determined, the epsilon-pole structure of any one-loop amplitude is known.

We therefore begin by constructing a basis of the dimension five and six operators which are required for renormalising QCD amplitudes in six dimensions. These operators contain either two quark fields and three derivatives, such as $\mathcal{O}_1 \equiv i \bar \psi \slashed{D} \slashed{D} \slashed{D} \psi$, or are purely bosonic operators such as $\tr F^{\mu\nu} F_{\nu\rho} F^\rho{}_\mu$.\footnote{Recall that a field strength $F$ counts as two derivatives since $[D_\mu, D_\nu] = -i g F_{\mu\nu}$.} A full list of potential operators appears in table~\ref{tab:operators}. 

Since we are only concerned with poles of on-shell amplitudes, rather than of off-shell correlation functions, we need only study operators which lead to independent contributions to the $S$ matrix. It is a well known fact that operators which are related by the classical equations of motion of the theory lead to the same contribution to the $S$ matrix, to all orders of perturbation theory~\cite{Wise:1979at,Politzer:1980me,Arzt:1993gz,GrosseKnetter:1993td,Simma:1993ky}. Thus we may simplify the list of operators in table~\ref{tab:operators} using the equations of motion
\begin{align}
i \slashed{D} \psi &= m \psi , \\
D^\mu F^a_{\mu\nu} &= -g \bar \psi \gamma^\nu T^a \psi.
\end{align}
It is straightforward to see that many operators in table~\ref{tab:operators} are related to other operators in our Lagrangian. For example,
\begin{equation}
\mathcal{O}_1 \equiv i \bar \psi \slashed{D} \slashed{D} \slashed{D} \psi = - i m^2 \bar \psi \slashed{D} \psi,
\end{equation}
so that $\mathcal{O}_1$ does not lead to a new, independent counterterm. It may therefore be omitted.

\begin{table}
\begin{center}
\begin{tabular}{ p{0.3\textwidth} | p{0.3\textwidth} | p{0.3\textwidth} }
Quark fields & \centering{Operator} & Operator class name \\
\hline
\multirow{9}{*}{Two quarks} & \centering{$ i\bar \psi \slashed D \slashed D \slashed D \psi $} & \multirow{3}{*}{$[\bar \psi D^3 \psi]$} \\
& \centering{$ i \bar \psi \slashed D D^2  \psi $} &  \\
& \centering{$ i \bar \psi D^\mu \slashed D D_\mu \psi$} & \\ 
\cline{2-3}
& \centering{$ \bar \psi \gamma^\mu \gamma^\nu F_{\mu\nu} \slashed D \psi $} & \multirow{3}{*}{$[\bar \psi D F \psi]$} \\
& \centering{$ \bar \psi D^\mu F_{\mu\nu} \gamma^\nu \psi $} &  \\
& \centering{$ \bar \psi F_{\mu\nu} \gamma^\mu D^\nu  \psi $} &  \\
\cline{2-3}
& \centering{$ \bar \psi \slashed D \slashed D  \psi $} & \multirow{2}{*}{$[\bar \psi D^2 \psi]$} \\
& \centering{$ \bar \psi D^2  \psi $} & \\
\cline{2-3}
& \centering{$ i \bar \psi \gamma^\mu \gamma^\nu F_{\mu\nu}  \psi $} & $[\bar \psi F \psi]$ \\
\hline
\multirow{3}{*}{Zero quarks} & \centering{$ i \tr F^{\mu\nu} F_{\nu \rho} F^\rho{}_\mu $} & \\
& \centering{$ \tr F^{\mu\nu} D^2 F_{\mu \nu} $} & \\
& \centering{$ \tr (D^\mu F_{\mu\nu}) (D^\rho F_\rho{}^\nu) $} & 
\end{tabular}
  \caption{Table of potential higher dimension operators in the 6 dimensional QCD effective Lagrangian. We have ignored four quark operators, which are not relevant for $t\bar t+\text{gluons}$ scattering at this order, and operators related to those in our table by integration-by-parts or Hermitian conjugation. We have also imposed the parity symmetry of QCD.}
\label{tab:operators}
\end{center}
\end{table}

Our task now is to construct a basis of operators which are independent under the use of the equations of motion, integration by parts etc. To construct such a basis, we consider several categories of operators. Firstly, we will focus on operators containing two quark fields. We classify these operators further according to the powers of derivatives, or of derivatives and field strength insertions as shown in detail in table~\ref{tab:operators}. We will begin by examining operators containing the largest number of derivatives or field strengths, as the use of the equations of motion may reduce these operators to simpler operators containing fewer derivatives (or field strengths.)

Each of the derivatives contained in operators of type $[\bar \psi D^3 \psi]$ has one Lorentz index which we must contract using either metric tensors or gamma matrices. By making use of the equations of motion, we may ignore the options of contracting the left-most or right-most $D$ index against a gamma matrix---such a contraction would reduce to an operator with fewer derivatives which we will analyze below. We are left with the unique possibility $\bar \psi D^\mu \slashed{D} D_\mu \psi$. However, this operator is equivalent to a linear combination of operators of class $[\bar \psi D F \psi]$ and $[\bar \psi D^2 \psi]$ upon use of the equations of motion, since
\begin{align}
\bar \psi D^\mu \slashed{D} D_\mu \psi 
&= \bar \psi\left( -i m D^\mu D_\mu  - i g D^\mu \gamma^\nu F_{\mu\nu} \right) \psi .
\end{align}
Therefore, the class $[\bar \psi D^3 \psi]$ can be completely reduced to simpler operators. 

Next, consider the class $[\bar \psi D F \psi]$. In this case we again have three possible Lorentz indices which must be contracted against gamma matrices or metric tensors. We may ignore the possibility of contracting the Lorentz index of the covariant derivative against a gamma matrix because of the equations of motion. We are left with two potential operator structures: $\bar \psi D^\mu F_{\mu\nu} \gamma^\nu \psi$ and $\bar \psi F_{\mu\nu} D^\mu \gamma^\nu \psi$. But
\begin{align}
\bar \psi D^\mu F_{\mu\nu} \gamma^\nu \psi &= \bar \psi (-g \bar \psi \gamma_\nu \psi) \gamma^\nu \psi + \bar \psi F_{\mu\nu} D^\mu \gamma^\nu \psi,
\end{align}
using the Yang-Mills equation. Since we are only interested in processes with two quarks, we will systematically ignore four quark operators. Therefore, we may replace the operator $\bar \psi D^\mu F_{\mu\nu} \gamma^\nu \psi$ with $\bar \psi F_{\mu\nu} D^\mu \gamma^\nu \psi$. This is the only member of the class $[\bar \psi D F \psi]$ which is of interest to us.

We now turn to operator structures containing two quark fields but only one extra power of derivatives or gauge fields. Thus the available operator structures are $[\bar \psi DD \psi]$ and $[\bar \psi F \psi]$. Up to equations of motion, there is only one operator of the first type: $\bar \psi D^\mu D_\mu \psi$. However, this is a reducible operator:
\begin{align}
\bar \psi D^\mu D_\mu \psi
&= \bar \psi \slashed{D} \slashed{D} \psi - \frac{ig}{2} \bar \psi F_{\mu\nu} \gamma^\nu \gamma^\mu \psi.
\end{align}
Thus, up to equations of motion, we may reduce the $[\bar \psi DD \psi]$ class to the $[\bar \psi F \psi]$ class. Because of the antisymmetry of the field strength tensor, there is only one operator in the $[\bar \psi F \psi]$ class, namely $\bar \psi F_{\mu\nu} \gamma^\nu \gamma^\mu \psi$. 

Finally, we must consider operators containing no quark fields. There are three gauge invariant possibilities: $\tr F^{\mu\nu} F_{\nu \rho} F^{\rho}{}_\mu$, $\tr F_{\mu \nu} D^2 F^{\mu\nu}$, and $\tr (D^\mu F_{\mu\nu}) (D^\rho F_{\rho}{}^\nu)$. The last of these three operators is equivalent to a four quark operator using the Yang-Mills equation, and is therefore of no interest to us. Meanwhile, the second of the three is equivalent to the other two:
\begin{align}
\tr F_{\mu\nu} D^2 F^{\mu\nu}
&= -2 \tr (D^\mu F_{\mu\nu}) D_\alpha F^{\alpha\nu} - 2ig \tr F_{\nu\mu} F^\mu{}_\alpha F^{\alpha\nu}.
\end{align}
As a result, we may also ignore this operator, leaving only $\tr F^{\mu\nu} F_{\nu \rho} F^{\rho}{}_\mu$.

In summary, there are only three higher dimension operators that contribute to the on-shell amplitudes. We may therefore take the full QCD Lagrangian in six dimensions, at one loop order, to be
\begin{align}
  \mathcal{L}_{QCD}^{6} = \bar \psi (i \slashed D -m) \psi &-\frac12 \tr F_{\mu\nu} F^{\mu\nu}  
    + \frac i2 \sigma_1 \, g_s^3 m_t \, \bar{\psi} \gamma^\mu \gamma^\nu F_{\mu\nu} \psi \nonumber \\
    &+ i \sigma_2 \, g_s^3 \, \bar{\psi} F_{\mu\nu} \gamma^\mu D^\nu \psi
    + \frac{i}6 \gamma \, g_s^3 \, {\rm tr}\left( F^{\mu\nu}[F_{\mu\lambda},F_\nu{}^\lambda]\right).
  \label{eq:6doperators}
\end{align}
A selection of the resulting Feynman rules are listed in Appendix \ref{app:6dfrules}.

We adopt the point of view that $\sigma_1$, $\sigma_2$ and $\gamma$ are couterterms which remove the divergences in loop amplitudes. In addition there are the usual counterterms from the
dimension four vertices 
$t\tb g$ and $ggg$. We can compute the constants $\delta_{t\tb g}, \delta_{ggg}, \sigma_1,
\sigma_2$ and $\gamma$ from simple one-loop vertex graphs. For example, expanding the $t\tb g$ vertex to
$\mathcal{O}(g_s^3)$ leads to,
\begin{align}
  \usepic{1.3cm}{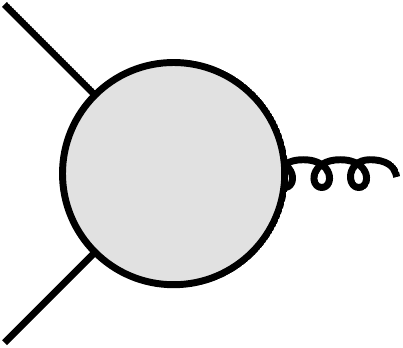}
  =
  &g_s \usepic{1.3cm}{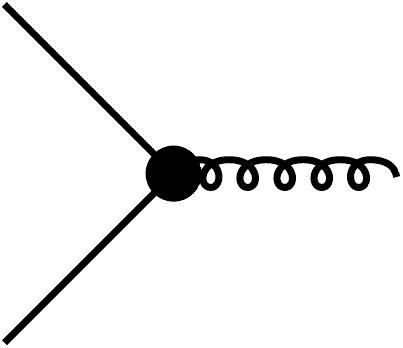}
  \nonumber\\
  + &g_s^3\Bigg(
  \usepic{1.3cm}{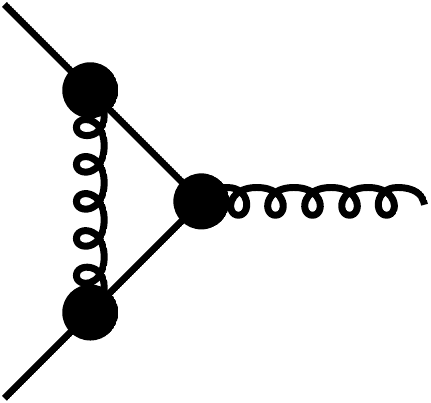}
  + \usepic{1.3cm}{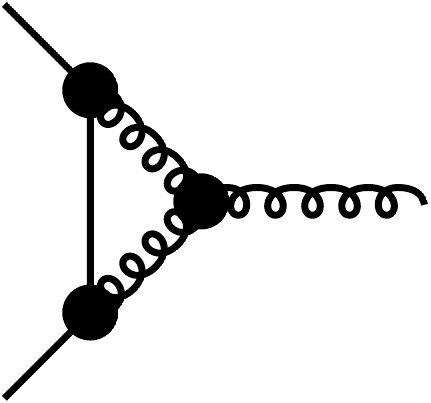}
  + \usepic{1.3cm}{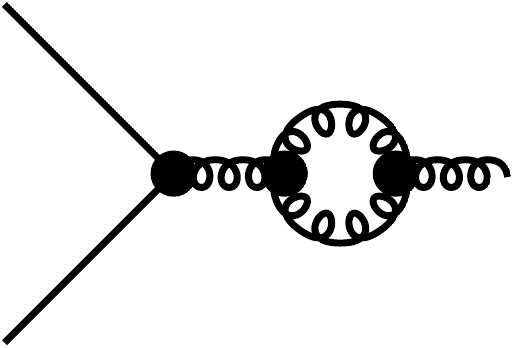}
  + \usepic{1.3cm}{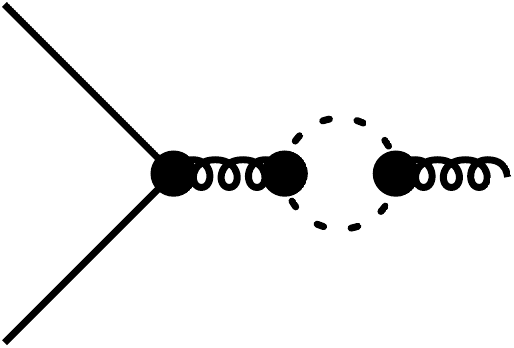} \nonumber\\&
  + \delta_{t\tb g} \usepic{1.3cm}{figures/qqg_tree.pdf}
  + \sigma_1 \usepic{1.3cm}{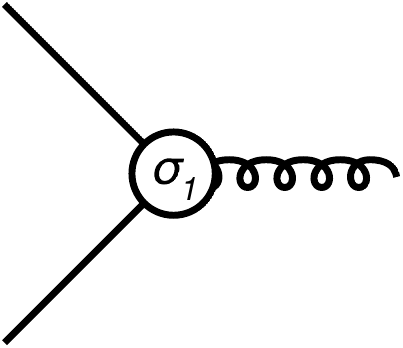}
  + \sigma_2 \usepic{1.3cm}{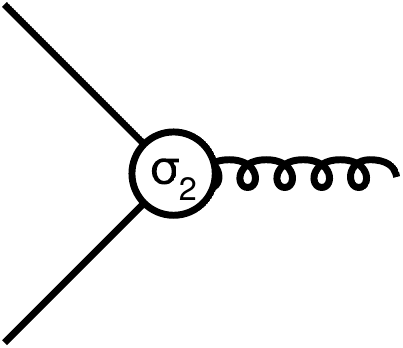}
  \Bigg)
  \nonumber\\
  + &\mathcal{O}(g_s^5).
\end{align}
Renormalising this correlation function off-shell would require the inclusion of all possible counterterms (before use of the equations of motion.) For us, it 
is simpler to compute the on-shell three point vertex, in which case all infinities can be absorbed in our effective Lagrangian, equation~\eqref{eq:6doperators}. This presents a minor problem since the three point
vertex is not well defined for real momenta. The computation may be performed using complex external
kinematics or alternatively performed with the gluon taken off-shell and the constants extracted by
taking the on-shell limit $p^2\to 0$ at the end of the computation. We find this amplitude is UV finite in
$6-2\eps$ dimensions for the values:
\begin{align}
  \delta_{t\tb g} &= \frac{m_t^2}{24 (4\pi)^3 \eps} C_F (3 d_s+2) ,\label{eq:deltattg}\\
  \sigma_1 &= -\frac{1}{12 (4\pi)^3 \eps} \left(C_A (d_s-5) - \frac{C_F}{2} (3 d_s - 14) \right),
  \label{eq:sigma1}
\end{align}
where $C_F = \frac{N_c^2-1}{2 N_c}$ and $C_A = N_c$. A similar computation for the three gluon vertex,
\begin{align}
  \usepic{1.3cm}{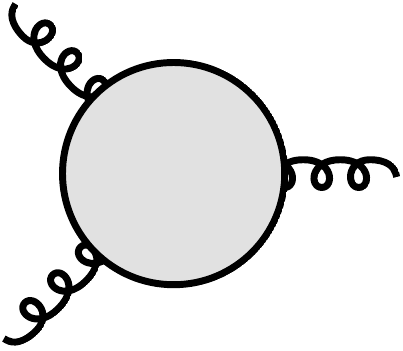}
  =
  &g_s \usepic{1.3cm}{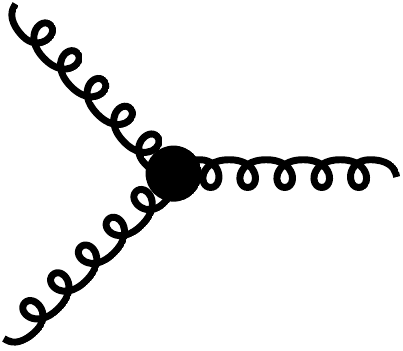}
  \nonumber\\
  + &g_s^3\Bigg(
    \usepic{1.3cm}{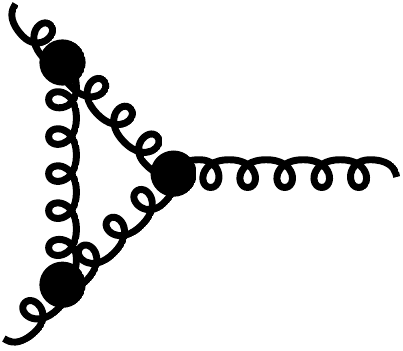}
  + \usepic{1.3cm}{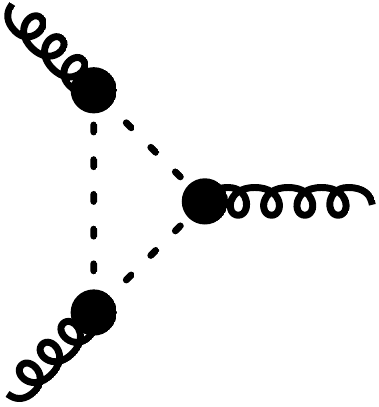}
  + \usepic{1.3cm}{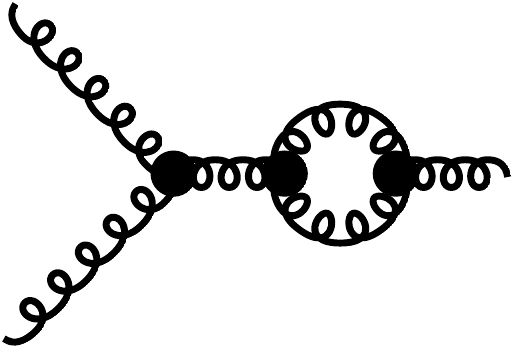}
  + \usepic{1.3cm}{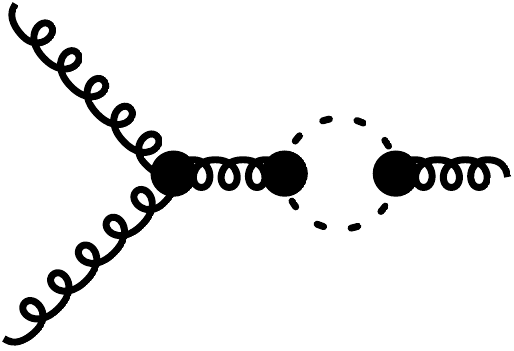}\nonumber\\&
  + \delta_{ggg} \usepic{1.3cm}{figures/ggg_tree.pdf}
  + \gamma \usepic{1.3cm}{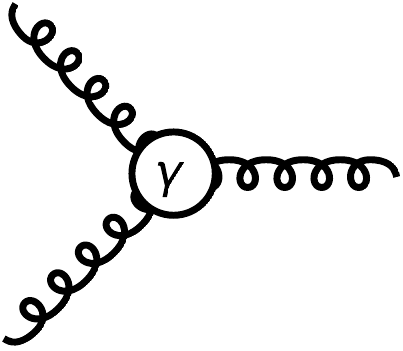}
  \Bigg)
  \nonumber\\
  + &\mathcal{O}(g_s^5),
\end{align}
results in
\begin{align}
  \delta_{ggg} &= 0 \label{eq:deltaggg} , \\
  \gamma &= \frac{1}{12 (4\pi)^3 \eps} C_A \frac{(d_s-2)}{5}. \label{eq:gamma}
\end{align}

\subsection{Fixing $c_1$ by matching poles in $6-2\eps$ dimensions \label{sec:tadpole}}

\begin{figure}[t]
  \begin{center}
     \begin{align*}
       A^{[\sigma_1]}(1_t,2,3,4_\tb) &=
            \usepic{2cm}{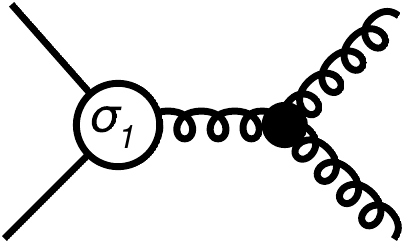}
          + \usepic{1.1cm}{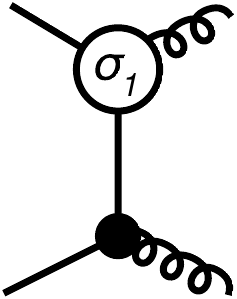}
          + \usepic{1.1cm}{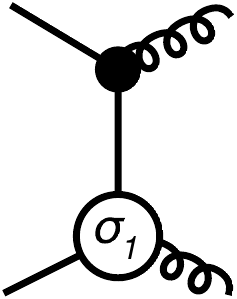}
          + \usepic{1.2cm}{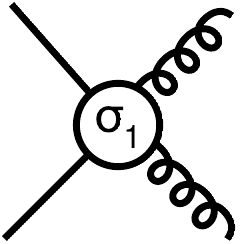} \\
       A^{[\gamma]}(1_t,2,3,4_\tb) &=
            \usepic{2cm}{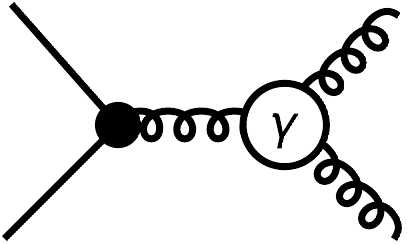}
     \end{align*}
  \end{center}
  \caption{The Feynman diagrams contributing to the tree-level amplitudes appearing the pole structure
  of the the one-loop $t\tb gg$ amplitudes in $6-2\eps$ dimensions. Solid vertices correspond to the usual QCD interactions
  while the open vertices are those resulting from the corresponding dimension six operators in
  $\mathcal{L}_{QCD}^{6}$ of eq. \eqref{eq:6doperators}.}
  \label{fig:6dttggtree}
\end{figure}

We finally apply this knowledge of the universal epsilon poles in six dimensions to determine
the remaining unknown coefficient, $c_1$ in equation~\eqref{eq:1l6dbasisshift}. The six-dimensional
leading colour partial amplitude $A_{4;1}^{(1),6-2\eps}(1_t,2,3,4_\tb)$ can be decomposed into gauge invariant
primitives
\begin{align}
  A_{4;1}^{(1),6-2\eps}(1_t,2,3,4_\tb) &=
       N_c A^{[L],6-2\eps}(1_t,2,3,4_\tb)
    - \frac{1}{N_c} A^{[R],6-2\eps}(1_t,2,3,4_\tb) ,
\end{align}
precisely as in four dimensions (we ignore fermion loops as they present no technical difficulties.)
Because the epsilon-poles are universal, we know that the poles of this amplitude are
\begin{align}
  A_{4;1}^{(1),6-2\eps}(1_t,2,3,4_\tb) &=
     g_s^4 \Bigg(
         2 \delta_{t\tb g} A^{(0)} (1_t,2,3,4_\tb)
        + \sigma_1 A^{[\sigma_1]} (1_t,2,3,4_\tb)\nonumber\\&
        + \sigma_2 A^{[\sigma_2]} (1_t,2,3,4_\tb)
        + \gamma A^{[\gamma]} (1_t,2,3,4_\tb)
      \Bigg)+ \mathcal{O}(\eps^0),
\end{align}
where the tree-type amplitudes $A^{[\sigma_1]} (1_t,2,3,4_\tb)$, $A^{[\sigma_2]} (1_t,2,3,4_\tb)$ and $A^{[\gamma]} (1_t,2,3,4_\tb)$ are associated with
the three higher dimension operators in the effective 6d QCD Lagrangian, equation~\eqref{eq:6doperators}. They are explicitly defined by the diagrams shown in figure \ref{fig:6dttggtree}. In a similar fashion to the vertex
computation we find that $A^{[\sigma_2]} (1_t,2,3,4_\tb) = 0$. By collecting in powers of
$N_c$, and inserting the known expressions for $\delta_{t\tb g}$, $\sigma_1$ and $\gamma$ given in equations \eqref{eq:deltattg}, \eqref{eq:sigma1} and \eqref{eq:gamma} we
find,
\begin{align}
  A^{[L],6-2\eps}(1_t,2,3,4_\tb) = \frac{g_s^4}{48(4\pi)^3 \eps}\Bigg(
  & 2 (3 d_s+2) m_t^2 A^{(0)} (1_t,2,3,4_\tb)
  + \frac{4(d_s-2)}{5} A^{[\gamma]} (1_t,2,3,4_\tb) \nonumber\\&
  - (d_s-6) A^{[\sigma_1]} (1_t,2,3,4_\tb)
  \Bigg) + \mathcal{O}(\eps^0)
  \label{eq:L6dpoles}
\end{align}
for the left-moving ordering and
\begin{align}
  A^{[R],6-2\eps}(1_t,2,3,4_\tb) = \frac{g_s^4}{48 (4\pi)^3 \eps}\Bigg(
  & 2 (3 d_s+2) m_t^2 A^{(0)} (1_t,2,3,4_\tb) \nonumber\\&
  + (3 d_s-14) A^{[\sigma_1]} (1_t,2,3,4_\tb)
  \Bigg) + \mathcal{O}(\eps^0)
  \label{eq:R6dpoles}
\end{align}
for the right-moving case.

The tree amplitudes $A^{[\sigma_1]} (1_t,2,3,4_\tb)$ and $A^{[\gamma]} (1_t,2,3,4_\tb)$ are easily determined by calculating the diagrams in figure \ref{fig:6dttggtree}. Written in terms of four dimensional spinor products, the independent helicity amplitudes are
\begin{align}
  -i&\spA{\n1}{\f1}\spA{\n4}{\f4} A^{[\sigma_1]}(1_t^+,2^+,3^+,4_\tb^+) = \nonumber\\&
    \frac{-2 m_t (2 m_t^2 - 4\d{\pp1}{\pp2} - \s{23})\s{23}\spA{\n{1}}{\p{2}}\spA{\n{4}}{\p{3}}}{\spA{\p{2}}{\p{3}}^3}
  + \frac{2 m_t (m_t^2 - 2\d{\pp1}{\pp2})\s{23}\spA{\n{1}}{\n{4}}}{m_t\spA{\p{2}}{\p{3}}^2}
  \nonumber\\&
  - \frac{m_t (m_t^2 - 2\d{\pp1}{\pp2})\s{23}\spA{\n{1}}{\p{3}}\spA{\n{4}}{\p{3}}\spAB{\p{2}}{\p{1}}{\p{3}}}{\d{\pp1}{\pp2}\spA{\p{2}}{\p{3}}^3}
  + \frac{m_t (m_t^2 - 2\d{\pp1}{\pp2})\s{23}\spA{\n{1}}{\p{2}}\spA{\n{4}}{\p{2}}\spAB{\p{3}}{\p{1}}{\p{2}}}{\d{\pp1}{\pp2}\spA{\p{2}}{\p{3}}^3},
  \\
  -i&\spA{\n1}{\f1}\spA{\n4}{\f4} A^{[\sigma_1]}(1_t^+,2^+,3^-,4_\tb^+) =
    \frac{(-4m_t(\d{\pp1}{\pp2})^2 + m_t^2\s{23} - 2\d{\pp1}{\pp2}\s{23})\spA{\n{1}}{\p{3}}\spA{\n{4}}{\p{3}}\spAB{\p{3}}{\p{1}}{\p{2}}}{\d{\pp1}{\pp2}\s{23}\spA{\p{2}}{\p{3}}}
  \nonumber\\&
  - \frac{2m_t\spA{\n{1}}{\n{4}}\spAB{\p{3}}{\p{1}}{\p{2}}^2}{m_t\s{23}}
  + \frac{m_t(4\d{\pp1}{\pp2} + \s{23})\spA{\n{1}}{\p{2}}\spA{\n{4}}{\p{3}}\spAB{\p{3}}{\p{1}}{\p{2}}^2}{\d{\pp1}{\pp2}\s{23}\spA{\p{2}}{\p{3}}}
  - \frac{m_t\spA{\n{1}}{\p{2}}\spA{\n{4}}{\p{2}}\spAB{\p{3}}{\p{1}}{\p{2}}^3}{\d{\pp1}{\pp2}\s{23}\spA{\p{2}}{\p{3}}},
  \\
  -i&\spA{\n1}{\f1}\spA{\n4}{\f4} A^{[\gamma]}(1_t^+,2^+,3^+,4_\tb^+) =
    \frac{m_t\s{23}^2\spA{\n{1}}{\p{2}}\spA{\n{4}}{\p{3}}}{2\spA{\p{2}}{\p{3}}^3}
  + \frac{m_t\d{\pp1}{\pp2}\s{23}\spA{\n{1}}{\n{4}}}{\spA{\p{2}}{\p{3}}^2} , \\
  -i&\spA{\n1}{\f1}\spA{\n4}{\f4} A^{[\gamma]}(1_t^+,2^+,3^-,4_\tb^+) = 0 , \\
  -i&\spA{\n1}{\f1}\spA{\n4}{\f4} A^{[\sigma_2]}(1_t^+,2^+,3^+,4_\tb^+) = 0 ,\\
  -i&\spA{\n1}{\f1}\spA{\n4}{\f4} A^{[\sigma_2]}(1_t^+,2^+,3^-,4_\tb^+) = 0.
\end{align}
We note that the amplitudes of the $\sigma_2$ operator vanish in the cases we have considered. 
We still include it in our analysis since the operator remains in the Lagrangian after using the equations of motion despite not
playing a role for the amplitudes in this paper.

The final step necessary to determine the tadpole coefficient is to evaluate the poles of the basis integrals
of the one-loop amplitude in $6-2\eps$ dimensions. We find
\begin{align}
  I^{6-2\eps}_1[1](m^2) &= \frac{-i m^4}{2(4\pi)^3\eps} + \mathcal{O}(\eps^0) \\
  I^{6-2\eps}_2[1](P^2, m_1^2, m_2^2) &= i \frac{P^2-3(m_1^2+m_2^2)}{6(4\pi)^3\eps} + \mathcal{O}(\eps^0) \\
  I^{6-2\eps}_2[\mu^2](P^2, m_1^2, m_2^2) &= i \frac{P^4-5P^2(m_1^2+m_2^2)+10\left((m_1^2+m_2^2)^2-m_1^2
  m_2^2\right)}{60(4\pi)^3\eps} + \mathcal{O}(\eps^0),
  \label{eq:6dintegralsbub}
\end{align}
\begin{align}
  I^{6-2\eps}_3[1] &= \frac{-i}{2(4\pi)^3\eps} + \mathcal{O}(\eps^0) \\
  I^{6-2\eps}_3[\mu^2](P_1^2, P_2^2, P_3^2, m_1^2,m_2^2,m_3^2) &=
  -i\frac{P_1^2+P_2^2+P_3^2-4(m_1^2+m_2^2+m_3^2)}{24(4\pi)^3\eps} + \mathcal{O}(\eps^0),
  \label{eq:6dintegralsbtri}
\end{align}
\begin{align}
  I^{6-2\eps}_4[1] &= \mathcal{O}(\eps^0), \\
  I^{6-2\eps}_4[\mu^2] &= \frac{i}{6(4\pi)^3\eps} + \mathcal{O}(\eps^0).
  \label{eq:6dintegralsbox}
\end{align}
We do not list the formulae for box integrals in 10 dimensions ($\mu^4$) since they do not appear in
amplitudes with a fermion pair and any number of gluons. The formulae are easy to derive using
the dimensional recurrence relation implemented in \textsc{LiteRed} \cite{Lee:2012cn} in any case.

The only unknowns in equations \eqref{eq:L6dpoles} and \eqref{eq:R6dpoles} are then the left- and right-moving tadpole coefficients $c_1$, allowing a
direct determination of these rational functions. The results are somewhat lengthy formulae which are explicitly derived in the \textsc{Mathematica} workbook. We have checked that this procedure matches the expected result by comparing with the previous computation of reference~\cite{Badger:2011yu}.

\section{Conclusions}\label{sec:conclusions}

In this paper we have explored a new technique for the computation of one-loop
amplitudes with massive fermions. Our methods are designed to be compatible with on-shell generalised unitarity.

The six-dimensional spinor-helicity scheme proved to be an efficient way to describe the tree-level
input into the $d$-dimensional generalised unitarity method. Divergent wavefunction cuts were
avoided, and the remaining ambiguities in the amplitudes were fixed by matching to the universal physical pole structure.
The $4-2\eps$ pole structure of Catani et al.~\cite{Catani:2000ef} is sufficient to constrain all remaining
logarithms in the fermion mass while additional information is needed to fix the remaining finite
corrections connected to tadpole integrals. We obtained this second constraint by allowing the loop
momenta in our integrals to be defined in a higher dimension spacetime, and imposing the universality
of ultraviolet divergences in this higher dimensional quantum field theory. Since six is the next even dimension
above four it was natural to study QCD as an effective theory in $6-2\eps$ dimensions.
We used the on-shell equations of motion to find a minimal set of additional dimension 
six operators in this theory, and computed the required counterterms essentially following the textbooks. 
We applied our method to a variety of simple cases and validated it on helicity amplitudes for top quark pair production.

The methods we used in this paper are flexible, and it is clear that they apply more generally than to $gg\to t\tb$ scattering. 
It would be interesting to work out the extension to more general cases with multiple fermions and multiple masses, as well as 
to higher loops.
In the presence of more fermions, four quark operators would need to be included in the effective Lagrangian,
while at higher loops one would need to consider operators of mass-dimension greater than six.

Since this method can compute amplitudes with fewer cuts than other known approaches it has the
potential to optimise existing numerical and analytical approaches. However, since the main
computational bottleneck in most phenomenological collider studies at NLO lies in the integration
over the unresolved phase-space, the technique is probably best suited to find compact analytic
expressions where the improvement in stability and speed over existing numerical approaches is
particularly beneficial.

Perhaps a more interesting direction would be to look into the implications of the higher dimensional
pole structure on the spurious singularities appearing in integral reductions. As a
result of matching to a tree-level computation with an effective Lagrangian, we find non-trivial
relations between the $d$-dimensional integral coefficients in which all spurious poles cancel.
These cancellations had to occur, since the effective theory contains only local operators. This information could
be useful in finding compact and stable representations of one-loop amplitudes.

\begin{acknowledgments}
  We would like to thank Fabrizio Caola, Richard Ball, Einan Gardi and Kirill Melnikov for useful discussions. SB
  is supported by an STFC Rutherford Fellowship ST/L004925/1 and CBH is supported by Rutherford
  Grant ST/M004104/1. DOC would like to thank the Urri\dh akv\'isl 25 Foundation for hospitality during
  completion of this work. DOC is an IPPP associate, and thanks the IPPP for on-going support as well as for hospitality during this work.
  This research was supported in part by the European Union through the ERC Advanced Grant MC@NNLO (340983).
\end{acknowledgments}

\appendix

\section{Conventions and spinor construction}\label{app:matrices}

We use the mostly minus metrics
\begin{align}
\eta^{\mu \nu} &= \text{diag}\{1,-1,-1,-1\}, \nonumber \\
\eta^{MN} &= \text{diag}\{1,-1,-1,-1,-1,-1\},
\end{align}
where lower-case Greek letters are used for four dimensional Lorentz indices and upper-case for six dimensions.

We use the following Pauli matrices:
\begin{align}
\sigma^1_{\alpha \dot{\alpha}} = \begin{pmatrix}
0 & -1 \\
-1 & 0
\end{pmatrix},\ \sigma^2_{\alpha \dot{\alpha}} = \begin{pmatrix}
0 & i \\
-i & 0
\end{pmatrix},\ \sigma^3_{\alpha \dot{\alpha}} = \begin{pmatrix}
-1 & 0 \\
0 & 1
\end{pmatrix},
\end{align}
and $(\tilde{\sigma}^\mu)^{\dot{\alpha}\alpha} = \epsilon^{\alpha \beta} \epsilon^{\dot{\alpha} \dot{\beta}}\sigma^\mu_{\beta \dot{\beta}}$ with $\epsilon_{12} = 1$. Using the Pauli matrices we define the $\Sigma$-matrices
\begin{align}
\Sigma^0 &= i \sigma^1 \times \sigma^2&\tilde{\Sigma}^0 &= -i \sigma^1 \times \sigma^2 \nonumber \\
\Sigma^1 &= i \sigma^2 \times \sigma^3&\tilde{\Sigma}^1 &= i \sigma^2 \times \sigma^3 \nonumber \\
\Sigma^2 &= \sigma^2 \times \sigma^0&\tilde{\Sigma}^2 &= - \sigma^2 \times \sigma^0 \nonumber \\
\Sigma^3 &= -i \sigma^2 \times \sigma^1&\tilde{\Sigma}^3 &= -i \sigma^2 \times \sigma^1 \nonumber \\
\Sigma^4 &= - \sigma^3 \times \sigma^2&\tilde{\Sigma}^4 &= \sigma^3 \times \sigma^2 \nonumber \\
\Sigma^5 &= -i \sigma^0 \times \sigma^2&\tilde{\Sigma}^5 &= -i \sigma^0 \times \sigma^2,
\end{align}
which obey the Clifford algebra
\begin{align}
\Sigma^M \tilde{\Sigma}^N + \Sigma^N \tilde{\Sigma}^M = 2 \eta^{MN}.
\end{align}

\section{Tree-level amplitudes in six dimensions} \label{app:treelevel6d}

In this section we list the tree-level amplitudes in six dimensions used in our calculation.

\subsection{Three-point amplitudes}
\begin{itemize}
\item{$A^{(0)}(1_{q}, 2_{\bar{q}}, 3_{g})$}
\begin{align}
A^{(0)} (1^{a}_{q}, 2^{b}_{\bar{q}}, 3^{c\dot{c}}_{g} ) =  \frac{i}{s_{r3}} \spaa{1^{a}}{2^{b}}{3^{c}}{r^{x}} \spab{r_{x}}{3^{\dot{c}}}
\end{align}
where $r$ is a massless reference vector satisfying $s_{r3}\neq 0$.

\item{$A^{(0)}(1_{g}, 2_{g}, 3_{g})$}
\begin{align}
A^{(0)}(1_{a\dot{a}}, 2_{b\dot{b}}, 3_{c\dot{c}}) & = i\Gamma_{abc}\tilde{\Gamma}_{\dot{a}\dot{b}\dot{c}} \quad &
\Gamma_{abc} &= u_{1a}u_{2b}w_{3c}+u_{1a}w_{2b}u_{3c}+w_{1a}u_{2b}u_{3c} \\ &&
\tilde{\Gamma}_{\dot{a}\dot{b}\dot{c}} & = \tilde{u}_{1\dot{a}}\tilde{u}_{2\dot{a}}\tilde{w}_{3\dot{c}}+\tilde{u}_{1\dot{a}}\tilde{w}_{2\dot{a}}\tilde{u}_{3\dot{c}}+\tilde{w}_{1\dot{a}}\tilde{u}_{2\dot{a}}\tilde{u}_{3\dot{c}},
\end{align}
where the tensors $\Gamma$ and $\tilde{\Gamma}$ are written in terms of the $SU(2)$ spinors $u, \tilde{u}$ satisfying the following properties, defined on a cyclic order $\{ijk\}$,
\begin{align}
u_{ia}\tilde{u}_{j\dot{b}} & = \spab{i_{a}}{j_{\dot{b}}}, &  u_{ja}\tilde{u}_{i\dot{b}} & = -\spab{j_{a}}{i_{\dot{b}}}  ,
\end{align}
and $w, \tilde{w}$ are the inverse of the $u, \tilde{u}$
\begin{align}
\epsilon_{ab} &=  u_{a}w_{b} - u_{b}w_{a},     &    \epsilon_{\dot{a}\dot{b}} & = \tilde{u}_{\dot{a}} \tilde{w}_{\dot{b}} -  \tilde{u}_{\dot{b}} \tilde{w}_{\dot{a}} ,
\end{align}
for which we impose momentum conservation
\begin{align}
0 & = \tilde{w}_{1\dot{a}}\tilde{\lambda}^{\dot{a}}_{1A} +\tilde{w}_{2\dot{a}}\tilde{\lambda}^{\dot{a}}_{2A}+\tilde{w}_{3\dot{a}}\tilde{\lambda}^{\dot{a}}_{3A}.
\end{align}

\item{$A^{(0)}(1_{\phi_{1,2}}, 2_{\phi_{1,2}}, 3_{g})$}
\begin{align}
A^{(0)}(1_{\phi_{1,2}}, 2_{\phi_{1,2}}, 3_{g}^{a\dot{a}}) = \frac{-i}{2s_{r3}} \la 3^{a}|(1-2)|r|3^{\dot{a}} ]
\end{align}
where $r$ is a massless reference vector satisfying $s_{r3}\neq 0$.

\item{$A^{(0)}(1_{\phi_{1}}, 2_{\bar{q}}, 3_{q})$}
\begin{align}
A^{(0)}(1_{\phi_{1}}, 2_{\bar{q}}^{a}, 3_{q}^{\dot{b}}) = \frac{i}{\sqrt{2}} \spab{1^{a}}{2^{\dot{b}}}\label{eq:amp3scalar1}.
\end{align}

\item{$A^{(0)}(1_{\phi_{2}}, 2_{\bar{q}}, 3_{q})$}
\begin{align}
A^{(0)}(1_{\phi_{1}}, 2_{\bar{q}}^{a}, 3_{q}^{\dot{b}}) = \frac{i}{\sqrt{2}} \spAB{1^a}{\gamma^5}{2^{\dot{b}}} \label{eq:amp3scalar5}.
\end{align}
\end{itemize}

\subsection{Four-point amplitudes}

\begin{itemize}
\item{$A^{(0)}(1_{g}, 2_{g}, 3_{g}, 4_{g})$}
\begin{align}
A^{(0)} (1_{a\dot{a}}, 2_{b\dot{b}}, 3_{c\dot{c}}, 4_{d \dot{d}} ) = \frac{-i}{s_{12}s_{23}} \spaa{1_{a}}{2_{b}}{3_{c}}{4_{d}} [1_{\dot{a}}2_{\dot{b}}3_{\dot{c}}4_{\dot{d}}]
\end{align}

\item{$A(1_{q}, 2_{g}, 3_{g}, 4_{\bar{q}})$}
\begin{align}
A^{(0)}(1_{q,a}, 2_{b \dot{b}}, 3_{c \dot{c}}, 4_{\bar{q},d}) &= \frac{i}{2s_{12} s_{23}} \langle 1_a 2_b 3_c 4_d \rangle [ 1_{\dot{x}} 2_{\dot{b}} 3_{\dot{c}} 1^{\dot{x}}].
\end{align}

\item{$A^{(0)}(1_{g}, 2_{g}, 3_{\phi_{1,2}}, 4_{\phi_{1,2}})$}
\begin{align}
A^{(0)} (1_{a\dot{a}}, 2_{b\dot{b}}, 3, 4) = \frac{i}{4s_{12}s_{23}} \spaa{1_{a}}{2_{b}}{3_{x}}{3^{x}} [1_{\dot{a}}2_{\dot{b}}3_{\dot{x}}3^{\dot{x}}]
\end{align}

\end{itemize}

\section{Cut solutions in six dimensions}\label{app:cutsols}

In this section we give details on the solutions for the triple- and double-cuts in six dimensions.
We will describe the parametrisation used to get the solutions
without writing down any explicit expression for them. The implementation is given in the attached \textsc{Mathematica} notebook.  
Notice that all the cut solutions are rational functions of the kinematics and the free parameters and contain no square roots. 

\subsection{Triple cut}

We write the loop momentum $\ell_{i}^{\mu}$ in the following basis,
\begin{align}
\beta = \left\{
	v^{\mu}, 
	w^{\mu}, 
	\spAA{v^{1}}{\Sigma^{\mu}}{w_{1}}, 
	\spAA{v^{1}}{\Sigma^{\mu}}{w_{2}}, 
	\spAA{v^{2}}{\Sigma^{\mu}}{w_{1}},
	\spAA{v^{2}}{\Sigma^{\mu}}{w_{2}}
	\right\},
\label{eq:basiscuts}
\end{align}
where $v$ and $w$ are six dimensional massless momenta and use the parametrisation
\begin{align}
\ell_{i} = \beta \cdot \left\{y_{1}, y_{2}, y_{3}, y_{4}, \tau_{1}, \tau_{2} \right\}.
\end{align}
We impose the cut conditions
\begin{align}
S_{ijk} = 
\begin{cases}
\ell_{i}^{2}=\ell_{j}^{2}=\ell_{k}^{2}=0 \\
\ell_{i}^{(5)}=
\begin{cases}
0 & \quad \text{if $i$ gluon} \\
\pm m & \quad \text{if $i$ fermion}
\end{cases}
\end{cases}, &
\label{eq:3cutsyst}
\end{align}
where $\{ijk\}$ is the set of the three cut propagators and the sign of the mass component depends on the kinematic configuration.
This system of equations for $\ell_{i}$ only constrains 4 parameters so solving for the $y_{i}$'s, $\tau_{1}, \tau_{2}$ are left as free parameters.

\subsection{Double cut}

For the double cut solutions we use the basis in \eqref{eq:basiscuts} and use the following parametrisation
\begin{align}
\ell_{i} = \beta \cdot \left\{y_{1}, \tau_{1}, y_{2}, \tau_{2}, y_{3}, \tau_{3} \right\}.
\end{align}
The $y_{i}$'s are fixed by the double cut constraints 
\begin{align}
S_{ij} = 
\begin{cases}
\ell_{i}^{2}=\ell_{j}^{2}=0 \\
\ell_{i}^{(5)}=
\begin{cases}
0 & \quad \text{if $i$ gluon} \\
\pm m & \quad \text{if $i$ fermion}
\end{cases}
\end{cases} &,
\label{eq:2cutsyst}
\end{align}
where $\{ij\}$ is the set of the two cut propagators and the sign of the mass component depends on the kinematic configuration.
The parameters $\tau_{1}, \tau_{2}, \tau_{3}$ are unconstrained.

\section{Feynman Rules for the effective Lagrangian}
\label{app:6dfrules}

In this appendix we present selected Feynman rules for the six dimensional effective theory of interest to us, defined by the Lagrangian
\begin{align}
  \mathcal{L}_{QCD}^{6} = \bar \psi (i \slashed D -m) \psi &-\frac12 \tr F_{\mu\nu} F^{\mu\nu}  
    + \frac i2 \sigma_1 \, g_s^3 m_t \, \bar{\psi} \gamma^\mu \gamma^\nu F_{\mu\nu} \psi \nonumber \\
    &+ i \sigma_2 \, g_s^3 \, \bar{\psi} F_{\mu\nu} \gamma^\mu D^\nu \psi
    + \frac{i}6 \gamma \, g_s^3 \, {\rm tr}\left( F^{\mu\nu}[F_{\mu\lambda},F_\nu{}^\lambda]\right).
\end{align}
We further define
\begin{align}
F^a_{\mu\nu} &= \partial_\mu A^a_\nu - \partial_\nu A^a_\mu + g_s f^{abc} A^b_\mu A^c_\nu, \\
\sigma^{\mu\nu} &= \frac i2 \left( \gamma^\mu  \gamma^\nu - \gamma^\nu \gamma^\mu \right).
\end{align}
These rules were derived with the help of FeynCalc~\cite{Mertig:1990an,Shtabovenko:2016sxi} and
FeynRules~\cite{Christensen:2008py,Alloul:2013bka}. The vertices are colour ordered and all
momenta are considered to be out-going. We include the coupling constants here for clarity
though in the main text they are stripped off.

\begin{align}
  \usepic{2cm}{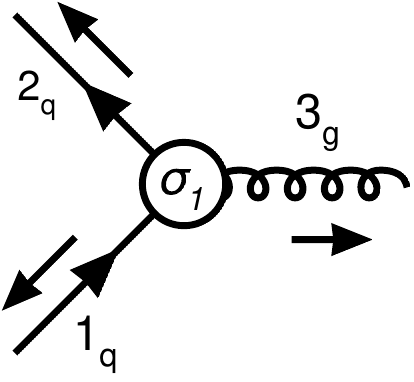}
  &=
     - g_s^3 \sigma_1 \, m_t \sigma^{\mu_3 \nu} p_{3\nu}
  \label{eq:vqqg6dsig1}\\
  \usepic{2cm}{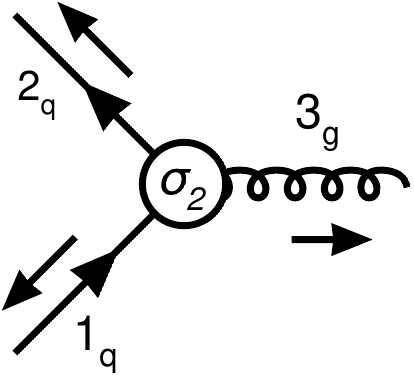}
  &=
     - i g_s^3 \sigma_2 \, \left( p_2^{\mu_3}\slashed{p}_3 - p_2\cdot p_3 \gamma^{\mu_3} \right)
  \label{eq:vqqg6dsig2}\\
  \usepic{2cm}{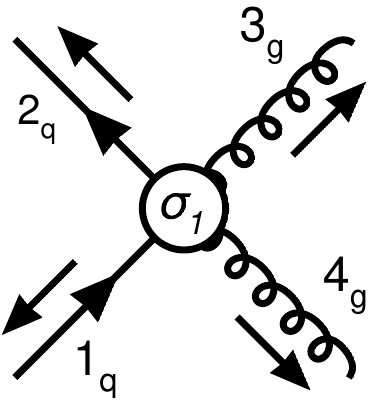}
  &=
     g_s^4 \sigma_1 \, m_t \sigma^{\mu_3 \mu_4}
  \label{eq:vqqgg6dsig1}\\
  \usepic{2cm}{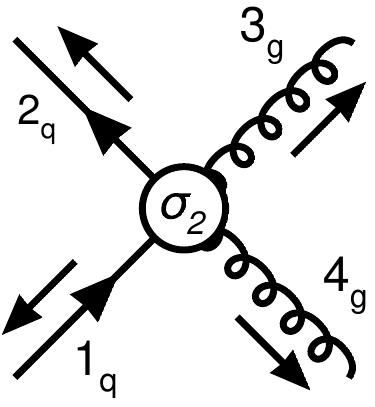}
  &=
     - i g_s^4 \sigma_2 \, \left( g^{\mu_3\mu_4}\slashed{p}_3 - \gamma^{\mu_4} p_1^{\mu_3} + \gamma^{\mu_3} (p_1^{\mu_4}-p_3^{\mu_4}) \right)
  \label{eq:vqqgg6dsig2}\\
  \usepic{2cm}{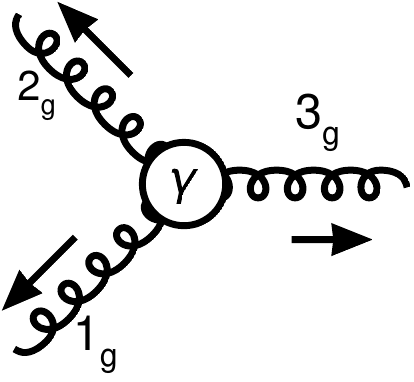}
  &=
  -\frac{i}{2} g_s^3 \gamma \, \Bigg( \nonumber\\&
    \hspace*{5mm}  g^{\mu_1\mu_2} \left( p_1\cdot p_3 \, p_2^{\mu_3} - p_2\cdot p_3 \, p_1^{\mu_3} \right)
      \nonumber\\&
    + g^{\mu_2\mu_3} \left( p_2\cdot p_1 \, p_3^{\mu_1} - p_3\cdot p_1 \, p_2^{\mu_1} \right)
    \nonumber\\&
    + g^{\mu_3\mu_1} \left( p_3\cdot p_2 \, p_1^{\mu_2} - p_1\cdot p_2 \, p_3^{\mu_2} \right)
    \nonumber\\&
    - p_3^{\mu_1} p_1^{\mu_2} p_2^{\mu_3} + p_2^{\mu_1} p_3^{\mu_2} p_1^{\mu_3}
    \Bigg).
  \label{eq:vggg6d}
\end{align}

\bibliographystyle{JHEP}
\bibliography{1l6dfermions}

\end{document}